\documentclass[aps,prc,reprint,amssymb,amsmath,superscriptaddress, floatfix,10pt]{revtex4-1}
\usepackage{mathptmx}
\usepackage{graphicx}
\usepackage{color}  
\makeindex
\newcommand{\textblue}{\textcolor[rgb]{0.00,0.07,1.00}}

\begin{document}

\title{Inner crust of neutron stars: Polymorphism and superconductivity in the liquid drop model}

\author{Dmitry Kobyakov}
\email{dmitry.kobyakov@appl.sci-nnov.ru}
\affiliation{Institute of Applied Physics of the Russian Academy of Sciences, 603950 Nizhny Novgorod, Russia}
\author{Xavier Vi\~nas}
\email{xavier@fqa.ub.edu}
\affiliation{Departament de F\'isica Qu\`antica i Astrof\'isica (FQA),
Universitat de Barcelona (UB), Mart\'i i Franqu\`es 1, E-08028 Barcelona, Spain}
\affiliation{Institut de Ci\`encies del Cosmos (ICCUB),
Universitat de Barcelona (UB), Mart\'i i Franqu\`es 1, E-08028 Barcelona, Spain}
\affiliation{Institut Menorqu\'i d'Estudis, Cam\'i des Castell 28, 07702 Ma\'o, Spain}

\begin{abstract}
Within the liquid drop model built up with the nuclear interaction parametrization Sk$\chi$450, which is based on the chiral effective field theory, we calculate numerically the internal energy density for each of nuclear pasta phases and for the uniform nuclear matter.
We provide quantitative arguments in favor of coexistence of various nuclear matter phases at a significant range of total pressure within the inner crust of neutron stars, a concept known as crystal polymorphism.
Specifically, we find that differences of the internal energy per baryon for various phases are typically less than the thermal energy per a freedom degree at temperature about $10^8$--$10^9$ K, which sets the energetic scale for thermal fluctuations of state of Fermi liquid from the ground state.
The nuclear energy contributions are described using the same parametrization Sk$\chi$450 for the bulk, plain surface and curvature terms.
We find that the introduction of the curvature correction changes the ground state in a relevant way.
This may be understood as a consequence of the corresponding change in size of the nucleus, which significantly modifies the phase transition densities.
Using the calculated structural parameters from liquid drop model, we explore the physical consequences of the expected Cooper pairing of protons in lasagna phase.
In this case, we find a crossover between the discreet layered and the three-dimensional anisotropic regimes of superconductivity. Additionally, we study the magnetic stress in lasagna accounting for a rotational lag between superfluid neutrons and the crystal lattice, which is believed to develop naturally in pulsars and magnetars.
Our results offer a preliminary insight into rich magnetic properties of the inner crust of neutron stars.
\end{abstract}

\maketitle

\section{Introduction}
\emph{Description of the problem.}
The physics of neutron star interiors represents a major challenge for both nuclear physics and astrophysics.
The ultimate question is: What is the structure of matter inside neutron star?
There are robust theoretical models for describing the structure of neutron stars up to densities of the order of 1-2 times the saturation
baryon number density $n_0$ in symmetric nuclear matter, which is of the order of $0.16$ fm$^{-3}$.
However, in the central part of the core of neutron stars the matter is expected to be at significantly higher densities.
At present, no reliable theoretical model exists to describe matter at such high density.
Fortunately, observations of neutron stars suggest physical phenomena that likely probe the global structure of neutron stars including the central part of the core.
Among those are type-II supernova events, which involve a collapsing stellar core with the relevant matter energy density \cite{BrownBetheBaym1982} and the gravitational wave observations \cite{OyamatsuIidaSotani2020}.
Another example is provided by giant magnetar flares and the corresponding quasi-periodic oscillations in the afterglows observed in X-ray astronomy \cite{TurollaZaneWatts2015}.
It is apparent from the energetic scale of these phenomena and from the corresponding basic physical picture of neutron star observable dynamics, that all these phenomena involve some kind of hydromagnetic waves spreading across the entire star.
Understanding the physics of hydromagnetic waves on a global stellar scale requires, first of all, to specify structure of the neutron star matter and properties of the electrical conduction of matter that determines its interaction with the magnetic field.

Since the pioneering quantitative studies of structure of the neutron star matter carried out by Baym, Bethe and Pethick \cite{BBP1971}, much work has been done in this direction.
It is convenient to categorize the studies according to the relevant length scale.
Here, we denote by the term \emph{microscopic structure} the nuclear matter structure at scales between $\sim 1$ fm and $\sim 50$ fm.
In contrast, the term \emph{macroscopic structure} is understood in this work as the length scales larger than $\sim 50$ fm, corresponding to thermodynamic subsystems which include many unit cells of the crystallized matter of the crust, assuming that for such subsystems the thermodynamic description (that is, for instance, involving temperature) is appropriate.

A notable step in microscopic studies was done in \cite{RavenhallEtAl1983,HashimotoEtAl1984}, unveiling the theoretical prediction that nuclear matter in the neutron star crust might exist in the form of nonsperical nuclei immersed in the pure neutron liquid.
Geometrical structure of such matter resembles the well-known varieties of ordinary pasta such as lasagna and spaghetti.
As a result, the term pasta phases has been adopted in nuclear astrophysics for the nonsperical nuclei in neutron star \cite{CaplanHorowitz2017}.
In the present work, we will focus on the inner crust matter, which is expected to host pasta phases.

\emph{Structure of the paper.}
In Section \ref{Sec_Overview} we present more details on the problem considered in this paper and provide a brief review of the relevant literature.
Liquid drop model used for description of the nuclear properties is explained in Section \ref{Sec_DescriptionLDM} with additional details presented in Appendices A-D.
In Section \ref{Sec_NumericalResultsLDM} we report the numerical solution to the equations of liquid drop model.
In Section \ref{Sec_SuperconductivityLDM} we propose a simple model for superconductivity incorporated into liquid drop model.
Using our numerical solution, in Section \ref{Sec_AnalysisNumericalResults} we analyze effects of temperature on the expected structure of the inner crust and evaluate some characteristic quantities of superconducting lasagna.
Section \ref{Sec_Astrophysical} is devoted to study of stress in a setting where a spherical layer of lasagna is subject to a pulsar spin lag between the charged particles and superfluid neutrons.

\section{Overview of studies of structure of pasta phases}
\label{Sec_Overview}
\emph{Studies of microscopic structure.}
Further studies of the microscopic structure of pasta phases have revealed that the energy differences per baryon between the different phases are quite small \cite{LattimerEtal1985}.
It has been found that the quantum corrections to the ground-state energy in neutron matter (the shell effects associated with unbound neutrons) are important and provide considerable uncertainty for understanding of the realistic structure of the crust \cite{BulgacMagierski2001,MagierskiHeenen2002}.
The effect originates both from structure at the microscopic scales where the precise geometry of Wigner-Seitz cell is defined, and from the macroscopic structure that contains information about the spatial distribution of structural lattice defects.

Within the mean-field quantum description based on the Schrodinger equation, the most technically sophisticated approach is provided by Hartree-Fock calculations (or Hartree-Fock-Bogoliubov equations if pairing is included).
There are many Hartree-Fock calculations of spherical nuclei since the seminal paper of Negele and Vautherin \cite{negele73} (see \cite{GrillMargueronSandulescu2011,ShelleyPastore2020} and also
\cite{mondal20} and references therein).
Lately, 3-dimensional mean-field calculations using non-relativistic and relativistic energy density functionals and classical and quantum molecular dynamics simulations have been performed \cite{NewtonEtal2022}.
Also, 1-dimensional time-dependent density functional theory accounting for band effects and superfluidity of neutrons has been established \cite{SekizawaEtal2022,YoshimuraSekizawa2024}
However, those calculations only provide partial information about the structure of the inner crust, because the problem of finding pasta phases with the minimum energy in the entire range of pressures corresponding to the inner crust has not been addressed in the full quantum approach.
Recently, the quantum effects have been studied in the spirit of Strutinsky energy theorem, as a perturbative correction to the quasiclassical Thomas-Fermi calculation in \cite{mondal20} and \cite{PearsonChamel2022}.
Thus, to date, no full quantum description to obtain the complete equation of state in the inner crust has been performed.

In studies of the structure, it is reasonable to start from liquid drop model \cite{BBP1971}, which allows to clearly separate
different contributions to thermodynamic quantities and paves the way to understanding various phenomena physically \cite{LattimerEtal1985}.
In liquid drop model, theoretical uncertainties stemming from the quasiclassical nature of the model and from the uncertainty of the input
nuclear energy functionals can be relatively easily evaluated and improved in a systematic way.

Microscopic description of the inner crust structure is also performed with the help of semiclassical approaches based on the Thomas-Fermi approximation (see \cite{sharma15} for more details) or through liquid drop model.
The latter does not involve the Schrodinger equation, but nevertheless is capable of a microscopic description on length scales of a single nucleus.
In addition, liquid drop model can be applied for inquiry regarding macroscopic, thermodynamic effects.

In this model, the properties of the nuclear surface, as for instance the surface tension and the behavior of the neutron and proton density mean-field profiles, play an essential role.
The surface tension has been always represented in approximate manner, based, for example, on the leptodermous expansion of the binding energy of nuclear systems \cite{myers69,myers74}.

In the leptodermous expansion, the orders of approximation correspond to the power of the moment in the integrand in Eqs. (\ref{eqA1}) and (\ref{eqA7}).
The leading approximation of liquid drop model is to calculate the surface tension associated with the planar surface (when the principal curvature is zero) with a sharp density profile of protons and use it for description of the nuclear clusters.
The approximate character of the surface tension in this case comes from the fact that in reality, the principal curvature of the surface of nuclear cluster is nonzero, the proton surface is smeared, and there is a thin surface layer of adsorbed neutrons at the surface of nuclear cluster.
In the next-to-leading approximation, the energy contribution due to a nonzero principal curvature, the proton surface
smearing and the neutrons adsorbed at the surface are included.
In this paper, we have only considered the curvature term, neglecting the two other contributions, whose impact is rather small.

Next-to-leading approximations of liquid drop model were considered in many recent liquid drop model calculations \cite{NakazatoIidaOyamatsu2011,newton13,LimHolt2017,carreau20a,carreau20b,dinh21a,dinh21b,
balliet21,shchechilin22}.
Parametrization of the planar surface tension in liquid drop model is often borrowed from the results of Lattimer, Pethick, Ravenhall and Lamb \cite{ravenhall83,LattimerEtal1985}, which were generalized later on by Lorentz and Pethick \cite{lorentz93}.
Interestingly, in many earlier calculations the curvature was neglected, while the planar surface contribution and the bulk contribution were taken from different parametrizations of the nuclear energy, raising questions about self-consistency.

In this paper, we improve on the earlier calculations and extract the planar and curvature contributions to the surface tension from a self-consistent extended Thomas-Fermi \cite{brack85} calculation performed using the same interaction that describes the bulk part in liquid drop model.
To this end we follow closely the method described in detail in Refs. \cite{Vinas1998,douchin00}, which is summarized in Appendix D.
The potential energy contribution to the energy density functional provided by Skyrme forces can be expressed in terms of the nucleon densities in uniform nuclear matter $n_n$ and $n_p$, and the square of their gradients, $({\bf \nabla}n_n)^2$ and $({\bf \nabla}n_p)^2$, which simulate the finite range of the nucleon-nucleon interaction.
In the extended Thomas-Fermi approach, the kinetic energy is also an energy density functional that is expanded in powers of $\hbar^2$.
In this work, we take only the two first terms of this expansion.
The leading order term corresponds to the bulk, which includes the Thomas-Fermi kinetic energy densities for neutrons and protons.
The next-to-leading order term contains the second order gradients coming from the Skyrme force and the Weizsacker correction to the kinetic energy density and effective masses, as well as a contribution of the semiclassical spin density, as it can be seen in Eqs. (\ref{eqA11}), (\ref{eqA12}) and (\ref{eqA13}).
We now turn to studies of the macroscopic structure.

\emph{Studies of macroscopic structure.}
The theoretical development has led to realization that the actual state of the crust is unlikely to be a system of stratified layers of pure pasta phases, but rather is likely to be a collection of monocrystals \cite{Jones1990,MagierskiHeenen2002,KobyakovPethick2015}.
The notion of polycrystalline matter that is constructed from monocrystals separated by planar structural defects, sounds plausible and is usually considered quite obvious \cite{Jones1990} by virtue of the analogy of the neutron star solid crust with known geological solids.
Still, it is expected that the mechanical nature of the pasta phases is much more complex than that of polycrystalline matter.
In fact, only the phases with spherical nuclei (or spherical bubbles, if exist) solidify into a solid state, while other pasta phases such as the spaghetti, lasagna and their bubble counterparts are very similar to liquid crystals \cite{PethickPotekhin1998,KobyakovPethick2018}.
Moreover, since the proton pairing is generally expected in nuclear matter near the nuclear saturation density \cite{BPP1969}, the protons inside the nonspherical nuclear clusters of the pasta phase should be superconducting \cite{Kobyakov2018,KobyakovPethick2018,ZhangPethick2021}.
Thus, the most striking physical difference with the usual liquid crystals is, perhaps, superconductivity of protons in the dense clusters of the pasta phases.

However, establishing further details in physical picture of the macroscopic structure is a complex task \cite{CaplanHorowitz2017}.
There are many factors which generate uncertainty within description of the crust structure on macroscopic scale.
For instance, it turns out that the stellar magnetic induction as moderate as $\sim 10^{15}$ G is important in calculations of the crust structure as soon as the proton pairing is taken into account \cite{Kobyakov2024}.
In this case, the proton pairing in the magnetic field leads to generation of the Meissner electric currents due to superconducting proton flow.
The latter tend to screen the magnetic field and to bunch its lines into the quantized magnetic vortices of proton fluid, while the electron magnetism can be neglected.
In lasagna, the ability of the material to develop the Meissner proton currents is clearly dependent on the relative angle between spatial orientation of lasagna structure and the direction of the magnetic field.
Some characteristic cases are demonstrated in panels (a) and (b) of Fig. 8 below.
The reason is that the superconducting electric currents might be impeded by the pure neutron matter, that fills the space between the nuclear clusters.
Therefore, energy of the interaction between the magnetic field and lasagna has an angular dependence.
This implies that there is an angular variation of the interaction energy.
The initial calculations in \cite{Kobyakov2024} have shown that this variation of the interaction energy is of the same order as the energy differences between various pasta phases, thereby pointing to an important role of a moderate strength magnetic field of $\sim 10^{15}$ G for investigation of the realistic structure of the inner crust.

Information regarding the superconducting gap energy and the microscopic structure of the pasta phases is crucial for the correct choice of the superconductivity model.
The latter is necessary for description of the magnetic properties of the pasta phases and, thus, is relevant for the astrophysical applications of the theory.
In lasagna, different superconductivity models correspond to qualitatively different physical situations, depending on the relation between strength of the proton pairing energy for two protons both from the same nuclear cluster or for two protons from the neighboring clusters.
From this viewpoint, the system is analogous to the terrestrial high-temperature superconductors.
The latter is a class of superconducting materials physically structured as a stack of thin electrically conducting layers separated by thin insulating layers \cite{KettersonSong1999}.
From experimental studies of the high-temperature superconductors, it is known that in case when a coherence length of the superconducting order parameter associated with quasi uniform matter inside the conducting layers is smaller than the distance between the edges of the conducting layers, then a discrete model of superconductivity is required.
A standard tool in this case is the Lawrence and Doniach model (see \cite{Clem1992} and references therein).
In the opposite case when the coherence length is larger than the distance between the edges of the conducting layers, the physical situation is qualitatively different and the appropriate description is given by the anisotropic Ginzburg-Landau (or, continuous) model \cite{KlemmEtal1975,DeutscherEntinWohlman1978,WangEtal2001}.

As a matter of fact, the continuous superconductivity model is generally valid at temperatures close to the superconducting critical temperature, if the superconducting transition temperature $T_c$ is smaller than the pasta crystallization temperature $T_m$.
This behavior is expected just below $T_c$, because in this case the coherence length $\xi$ becomes very large and easily exceeds the distance $d_L$ between the clusters.
However, the continuous model might be invalid in the low-temperature regime well below the superconducting transition temperature, if $\xi$ becomes smaller than $d_L$.

Finally, it is worth to note that the discreet superconductivity model was assumed in \cite{Kobyakov2018}.
One of the most interesting consequences for such assumption is that, in the absence of bridges or proton tunneling connecting the neighboring slabs in lasagna, the penetration depth of the magnetic field which is parallel to the slab surface, is infinite.
This, in turn, implies that under some symmetric conditions (see Fig. 10 below), the magnetic stress might well exceed the yielding stress of the crust material \cite{Kobyakov2018}.
In the further work, Zhang and Pethick \cite{ZhangPethick2021} have used a continuous model for superconductivity in lasagna
phase.

In this work, we will introduce a simple model of the superconductivity in lasagna and extract the density-dependent value of superconducting energy gap inside the nuclear cluster of lasagna from the calculations of \cite{LimHolt2021} for uniform nuclear matter.
As a result, we shall demonstrate that in defect-free lasagna, a crossover between the discreet and the continuous superconductivity regimes might take place at some relevant pressure.
Certainly, such approach is only a rather crude approximation, because (i) in lasagna, the superconducting gap might be anisotropic in the momentum space even if S-wave pairing for neutrons is assumed and (ii) matter inside the clusters of lasagna is not homogeneous \cite{EstalEtal2001,PankratovEtal2008,SchuckVinas2011,OkamotoEtal2013,FurtadoEtal2022,YoshimuraSekizawa2024}.
However, we think that this simplified approach is a reasonable starting point for understanding of physics for superconductivity effects in magnetic field.

\section{Description of liquid drop model}
\label{Sec_DescriptionLDM}
\subsection{Basic approximations}
In order to study quantitatively the microscopic structure of nuclear matter in the conditions of the inner crust of neutron stars, we first consider a theoretical model, in which the matter is assumed to be in its ground state at zero temperature.
This implies an ideal crystal lattice, which is conveniently described in terms of a minimal unit cell (or, Wigner-Seitz cell, or simply the unit cell).
We also assume that the electrons are uniform at the scale of a single unit cell, which is a good approximation to the realistic state of the electrons due to the fact that their many-body state is degenerate and ultrarelativistic.
The baryon density averaged at length scales much larger than size of the unit cell is denoted as $n_b$.
Locally inside the nuclear cluster, the matter is assumed to have a uniform density $n$ associated with baryons with a local number density of protons $n_p=xn$, $x$ being the proton fraction, and the local density of neutrons $n_{ni}=(1-x)n$.
The voids between the nuclei are filled with the dripped neutrons with a uniform density $n_{no}$.

We focus on the unit cell with volume $V_c$, which contains the total number of neutrons $N_n$ and protons $N_p$:
\begin{equation}\label{def_nb}
  n_b=\frac{N_p+N_n}{V_c}.
\end{equation}
A cell either contains a nucleus surrounded by the neutron liquid.
In case of the bubble phases, the unit cell is entirely filled with the nuclear matter except in its center, where there is a bubble filled with a pure neutron liquid.
Correspondingly, the types of unit cell are called pasta phases and are denoted as 1N (lasagna), 2N (spaghetti), 3N or, for the bubble phases, 1B (lasagna), 2B, 3B.

In case of 3N and 3B, the nuclear matter and the pure neutron liquid are separated by a spherical interface.
Correspondingly, in case of 2N and 2B the interface is a cylindrical surface with infinite length.
In 1N and 1B phases, the interface is planar surface with infinite area.
The unit cells have a finite lattice period along three spatial dimensions (for 3N and 3B), or along two dimensions (for 2N and 2B), or along only a single dimension (for 1N and 1B).

Notice that in 1N and 1B phases, the picture in terms of 1N is equivalent to the picture in terms of 1B with an appropriate change in variables, and thus, the results for 1N and 1B are combined and represented as a single case, which is denoted briefly as 1N (or, lasagna).
Still, it is important that the volume fraction for 1N corresponds to the ratio between the volume of the nucleus and the unit cell, while the volume fraction for 1B corresponds to the ratio between the volume of neutron-filled bubble and the unit cell.
Hence the variable $u$ used for the volume fraction in the phase of nuclear clusters is distinct from the variable $u^{\rm bub}$ used for the volume fraction in the phase of nuclear bubbles.
To keep the theory as clear as possible, in Appendices B-C we present the formulas for nuclei and for bubbles separately.
Finally, we note that we use the most symmetric approximation for the unit cell (such as the spherical unit cell in 3N and 3B phases), which leads to certain relation between the volume fraction and the nucleus radius, see Eqs. (\ref{def_u}) and (\ref{def_u_BUB}).

\subsection{Basic equations and scheme of numerical solution}
As it is explained in Appendix B, within the unit cell the total energy density, Eq. (\ref{wtot}), for 1N, 2N and 3N phases can be written as
a sum of the following pieces:
\begin{eqnarray}\label{wtotSect1}
w_{\mathrm{tot}}
 =w_{\mathrm{nuc}} + w_{\mathrm{p.surf}} + w_{\mathrm{C+L}} + w_{\mathrm{no}} + w_{\mathrm{e}} + w_{\mathrm{curv}},
\nonumber \\
\end{eqnarray}
where $w_{\mathrm{nuc}}$ is the bulk (uniform) energy density of the nuclear cluster.
Analogously, the energy density for bubble phases is described by Eq. (\ref{wtot_BUB}) and is explained in Appendix C.
We neglect a contribution from the neutron skin, a correction due to smearing of the nucleon densities and a correction due to pairing energy.

In the total energy density $w_{\mathrm{tot}}$, the terms  $w_{\mathrm{p.surf}}$ and $w_{\mathrm{curv}}$ are the nuclear surface contributions, where the former describes the \emph{planar surface} contribution and the latter describes the \emph{curvature} of the nucleus interface.
The contribution from the dripped neutron liquid is given by $w_{\mathrm{no}}$ and the kinetic energy of the electron background by $w_{\mathrm{e}}$. The Coulomb energy term $w_{\mathrm{C+L}}$, collects the self-energy of protons in the nuclear cluster and of electrons distributed in the WS cell as well as the proton-electron lattice energy.

The basic equations are a set of coupled algebraic equations, which can be obtained from the variational procedure, as specified in Appendices B-C.
Satisfaction of the variational equations implies the system state is in the local energy extremum.
With the mean baryon density $n_b$ given as an external parameter, the four unknowns are the proton fraction inside the nucleus $x$, the nucleus radius $r_N$, the baryon number density inside the nucleus $n$ and the neutron number density inside the pure neutron liquid $n_{no}$:
\begin{equation}\label{NvariablesPractMainText}
  \{x,\,r_N,\,n,\,n_{no}\}.
\end{equation}
Correspondingly, in the bubble phase the unknowns are the proton fraction outside the bubble $x$, the bubble radius $r_B$, the baryon number density outside the bubble $n$ and the neutron number density inside the bubble $n_{no}$.

The full set of basic equations contains four algebraic equations:
\begin{eqnarray}
  \label{var_1} && \mu_{e}=\mu_{ni} - \mu_{pi}, \\
  \label{var_2} && w_{\rm p.surf} + 2w_{\rm curv}=2w_{\rm C+L}, \\
  \label{var_3} && \mu_{ni}=\mu_{no}, \\
  && \label{var_4} P_i=P_o.
\end{eqnarray}
The quantities in Eqs. (\ref{var_1})-(\ref{var_4}) are defined for the nucleus phases in Eqs. (\ref{betaEquilCond_form}), (\ref{Muni}) and (\ref{Mupi});
(\ref{ws_2wCL}), (\ref{def_ws}), (\ref{def_wCoul}) and (\ref{def_wbend}); (\ref{continuityMu}), (\ref{Muni}) and (\ref{Muno}); (\ref{continuityP}),
(\ref{Pi}) and (\ref{Po}).
The corresponding quantities for the bubble phases are defined in Eqs. (\ref{betaEquilCond_form_BUB}), (\ref{Muni_BUB}) and (\ref{Mupi_BUB}); (\ref{ws_2wCL_BUB}), (\ref{def_ws_BUB}), (\ref{def_wCoul_BUB}) and (\ref{def_wbend_BUB}); (\ref{continuityMu_BUB}), (\ref{Muni_BUB}) and (\ref{Muno_BUB}); (\ref{continuityP_BUB}), (\ref{Pi_BUB}) and (\ref{Po_BUB}).
Notice that our definitions of the pressure inside the nucleus $P_i$ and of the chemical potential of neutrons inside the nucleus $\mu_{ni}$,
include both the bulk and the surface contributions, which differs from the notation used by Lim and Holt \cite{LimHolt2017}.

In order to solve Eqs. (\ref{var_1})-(\ref{var_4}), as a first step we choose the ranges of values of $n$ and $n_{no}$ and represent them as
uniform arrays of points.
For each pair of values of $n$ and $n_{no}$ from the chosen ranges we find from Eq. (\ref{var_1}) the equilibrium value of $x=x(n,n_{no})$.
In this step we express the volume fraction $u$ from Eqs. (\ref{def_nb}) and (\ref{def_nb_BUB}) as
\begin{equation}\label{u_expr}
  u=\frac{n_b-n_{no}}{n-n_{no}},\quad u^{\rm bub}=\frac{n-n_b}{n-n_{no}},
\end{equation}
which allows to find $r_N$ and $r_B$ using Eq. (\ref{var_2}).

In the next step we find the pairs of variables $(n,n_{no})$ that satisfy the rest two of the variational equations, Eqs. (\ref{var_3}) and
(\ref{var_4}).
As a result, we obtain two curves on the plane $(n,n_{no})$, which might intersect thereby determining the point of existence of the numerical solution.
Having determined the intersection point, we have found all the independent variables of the problem and thus, the solution is accomplished.

\section{Liquid drop model: Numerical results}
\label{Sec_NumericalResultsLDM}
\subsection{Verification of computational algorithm and choice of surface tension}
We start with a numerical verification of our algorithm and reproduce the earlier landmark results, in particular, the appearance of various pasta
phases and their transition densities as given in Table III of \cite{LimHolt2017}.

As it is described in the Appendix D, we have computed self-consistently the surface tension $\sigma_s(x)$ and the curvature correction
$\sigma_c(x)$ using the same nuclear interaction, Sk$\chi$450, as for calculating the bulk.
To obtain the main result of this paper, we use directly the numerical data given in Table \ref{table2} for $\sigma_s(x)$, $\sigma_c(x)$ and
their derivatives.
This is in contrast to earlier liquid drop model calculations, where different equations of state were used for describing the bulk matter but
with a surface tension given by an analytical expression depending on few parameters \cite{ravenhall83,LattimerEtal1985,lorentz93}.
For instance, Eq. (9) of \cite{LimHolt2017} suggests the following fit
\begin{equation}\label{def_sigmaLimHolt}
  \sigma_s^{\rm fit}(x) = \sigma_0\frac{2^{\alpha+1} + q}{(1-x)^{-\alpha} + q + x^{-\alpha}},
\end{equation}
where the numerical values of the parameters are:
\begin{equation}\label{FitParameters}
\sigma_0=1.186,\quad \alpha=3.4,\quad q=46.748.
\end{equation}
We have found that our numerical data on $\sigma_s(x)$ given in Table \ref{table2} can also be fitted by the function Eq. (\ref{def_sigmaLimHolt}) with the following coefficients:
\begin{equation}\label{FitParametersVinas}
\sigma_0=1.1110, \quad \alpha=3.5281, \quad q=39.3491.
\end{equation}

In order to check our numerical algorithm, we compute the inner crust structure using Eq. (\ref{def_sigmaLimHolt}) with the parameters given in
Eq. (\ref{FitParameters}) and discarding the curvature contribution.
For convenience, we use throughout the paper the scaled variable
\begin{equation}\label{def_eta}
  \eta\equiv\frac{n}{n_0}.
\end{equation}
We calculate the transition densities between different structures with numerical error bars and display the results in the third column of
Table I (marked as Case 2).
For comparison, we list the earlier results due to Lim and Holt \cite{LimHolt2017} in the second column of
Table I (marked as Case 1).
We observe a good agreement between Case 1 and Case 2 in Table I.
This provides an independent verification of our numerical algorithm.
We have also verified that two sets of numerical results (with $\sigma_c=0$) obtained from two different parameterizations given in Eqs. (\ref{FitParametersVinas}) and (\ref{FitParameters}), are not significantly different.

As a next step, we include both the planar surface contribution $\sigma_s(x)$ and the curvature contribution $\sigma_c(x)$, using the numerical
data given in Table \ref{table2} for $\sigma_s(x)$, $\sigma_c(x)$ and their derivatives.
The transitions between different configurations are listed in the fourth column of Table \ref{table3} (marked as Case 3).
A comparison between the results shown in the the second and fourth columns of Table \ref{table3} (marked as Case 1 and Case 3) suggests that, the inclusion of $\sigma_c(x)$ in this paper induces a qualitative difference for the ground state of the inner crust matter, as compared with the ground state predicted earlier in \cite{LimHolt2017}.

\begin{table}
\begin{tabular}{|c|c|c|c|}
  \hline
  & $\eta$ (Case 1) & $\eta$ (Case 2) & $\eta$ (Case 3) \\
  \hline
  3N-2N & 0.4061 & 0.4025$\pm$0.002 & 0.3068$\pm$0.002 \\
  \hline
  2N-1N & 0.4715 & 0.4664$\pm$0.002 & 0.3666$\pm$0.002  \\
  \hline
  1N-2B & 0.5361 & 0.5342$\pm$0.002 & --  \\
  \hline
  2B-3B & 0.5502 & 0.5501$\pm$0.002 & --  \\
  \hline
  3B-Uni. & 0.5624 & 0.5621$\pm$0.002 & --  \\
  \hline
  1N-Uni. & -- & -- &  0.5581$\pm$0.002 \\
  \hline
  $\eta_{\rm uni}^{*}$ & -- & 0.5508 &  0.5508 \\
  \hline
  $\eta_{\rm p}^{*}$ & -- & 0.5770$\pm$0.002 &  0.5771$\pm$0.002 \\
  \hline
\end{tabular}
\caption{\label{table3} The baryon number densities $\eta\equiv n_b/n_0$ at which the solution with the minimum energy per baryon changes its
symmetry type (between 1N, 2N, 3N, 2B, 3B phases), or $\eta_{\rm uni}^{*}$ at which the uniform nuclear matter is unstable with respect to
modulations of the nucleon densities, or $\eta_{\rm p}^{*}$ at which the proton drip occurs (see Fig. 5), in various cases.
\emph{Case 1}: Output data from Table III of \cite{LimHolt2017}.
\emph{Case 2}: Solutions to Eqs. (\ref{var_1})-(\ref{var_4}) with $\sigma_c=0$ and with $\sigma_s$ from Eqs. (\ref{def_sigmaLimHolt}) and
(\ref{FitParameters}).
\emph{Case 3}: Solutions to Eqs. (\ref{var_1})-(\ref{var_4}) with $\sigma_s$ and $\sigma_c$ from Table \ref{table2}.
The quantity $\eta_{\rm uni}^{*}$ is found from \cite{LimHolt2017} (see Eq. (30) and Table V in \cite{LimHolt2017}).
The quantity $\eta_{\rm p}^{*}$ is found from our solution for the workfunction $\Delta\mu_p$ shown in Fig. 5.
In this work we focus on Case 3, in which the inclusion of curvature energy leads to dramatic effects, specifically, to disappearance of the bubble phases.
}
\end{table}

\subsection{Energies of pairing, neutron skin and density smearing}
In the main numerical calculations of this paper, we have ignored energies of pairing, neutron skin and density smearing.
It appears from calculations by B. K. Sharma \cite{Sharma2024} that the neutron skin and density smearing do not significantly affect the pasta phase transition.
This is in contrast to the curvature correction, which notably influences the appearance of pasta phases, as can be seen from the results reported in Table I.

We have performed calculations of energies of the pasta phases with the pairing energy density included by adding $w_p$ from Eq. (\ref{def_w_pairing}) (and for the bubble phases, $w_p^{\rm bub}$ from Eq. (\ref{def_w_pairing_BUB}), correspondingly) to the total energy density $w_{\rm tot}$ given in Eq. (\ref{def_wtot}).
We have found that the pairing has no effect on the pasta phase transition densities within the actual precision of our computation.
The reason is that $w_p$ is small in magnitude and it does not influence the nucleus radius.

\subsection{Main numerical results: Solution with curvature}
In the main calculation of this paper, we have incorporated the curvature correction to the surface tension, which is a contribution to the energy due to the principal curvature of a thin nuclear surface without finite-width corrections.
As we have mentioned above, we use our self-consistent numerical results (as given in Table \ref{table2}) on the leading (planar surface) contribution $\sigma_s(x)$ and the next-to-leading (curvature) correction $\sigma_c(x)$, for the surface tension of a thin surface.
We represent these functions numerically with the help of spline interpolation -- in this way we avoid choosing any fitting parameters such as
in Eq. (\ref{def_sigmaLimHolt}).

By following the scheme of solution of the variational equations explained above, we find the four basic variables,
$\{n,x,r_N,n_{no}\}$ (and the corresponding quantities for the bubble phases) as functions of mean (averaged across the WS cell) baryon density $n_b$ scaled by the saturation density in symmetric nuclear matter $n_0$ and display them in Figs. 1--4.

Figures 1-4 show, correspondingly, the proton fraction $x$ inside the nucleus (or outside the bubble), the nucleus radius $r_N$ (or the scaled bubble radius, $r_B\times10^{-1}$), the local baryon density $n$ inside the nucleus (or outside the bubble) and the local number density $n_{no}$ of neutrons outside of the nuclear cluster (or inside the bubble).
\begin{figure}
\includegraphics[width=3.5in]{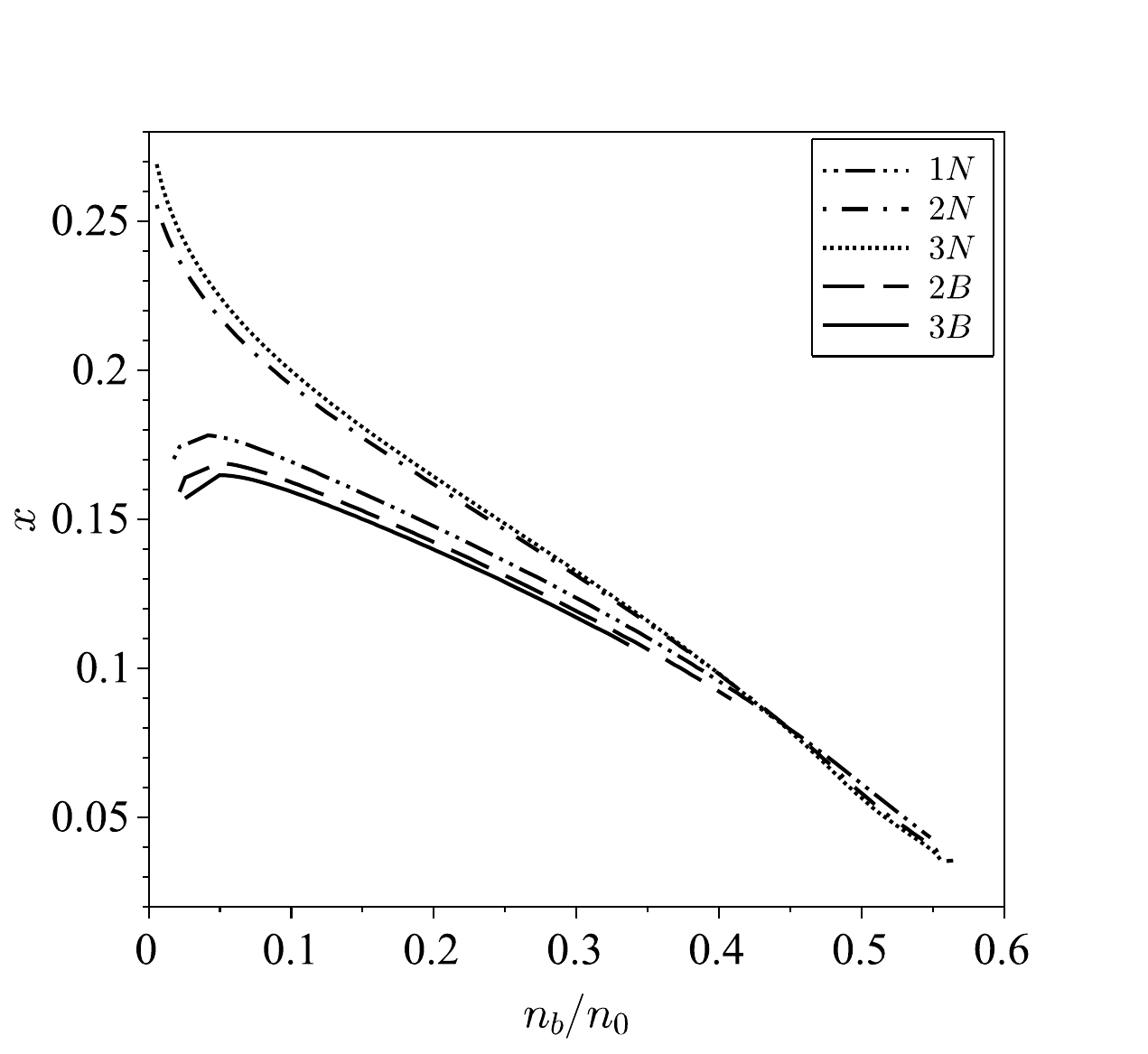}
\caption{Solution to the basic equations of equilibrium, Eqs. (\ref{var_1})-(\ref{var_4}): The proton fraction $x$ inside the nucleus,
Eq. (\ref{def_x}), and outside the bubble, Eq. (\ref{def_x_BUB}), as function of the mean baryon density $n_b$ (scaled by $n_0$).}
\end{figure}
\begin{figure}
\includegraphics[width=3.5in]{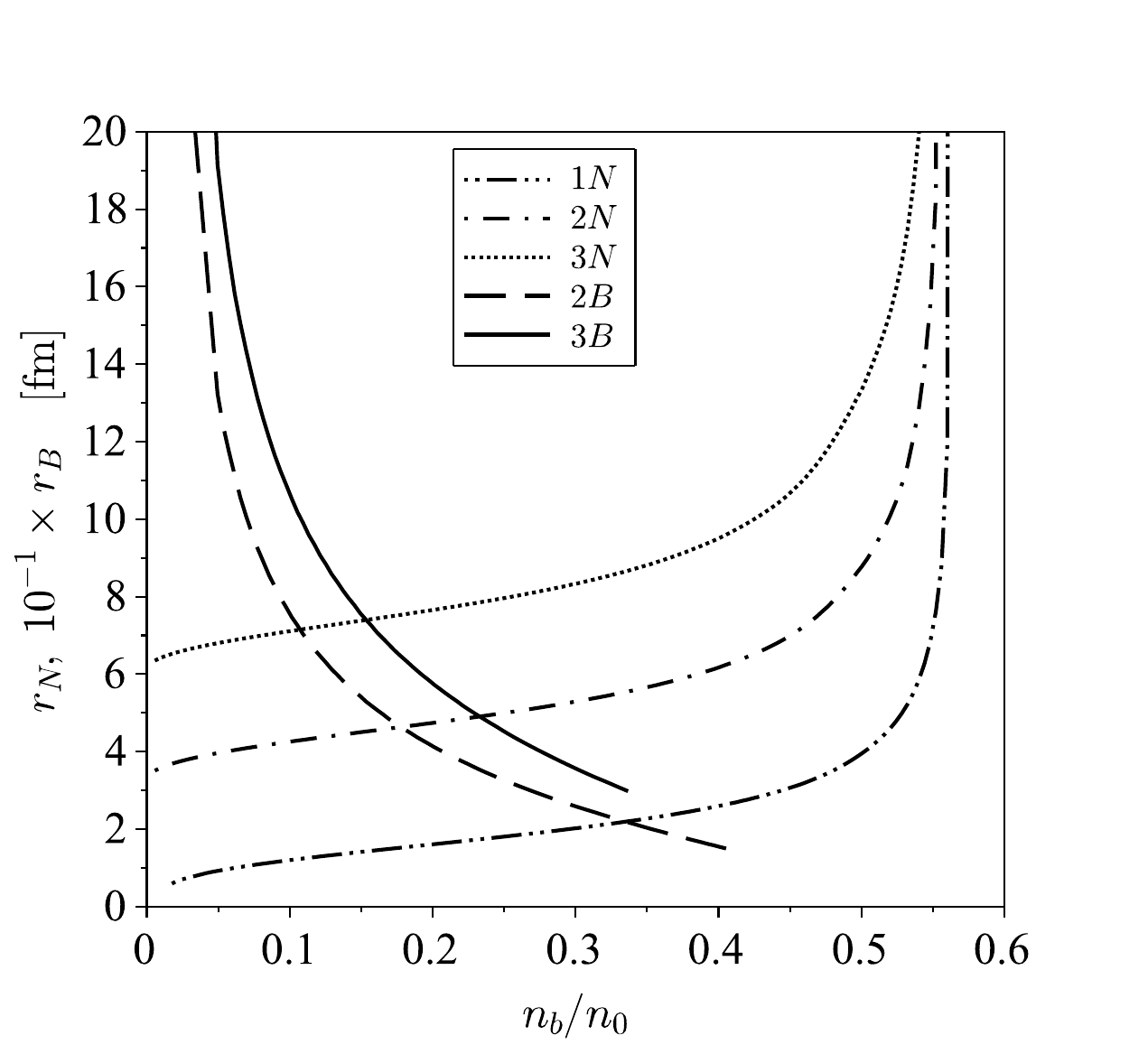}
\caption{Solution to the basic equations of equilibrium, Eqs. (\ref{var_1})-(\ref{var_4}): The nucleus radius $r_N$ and the scaled bubble radius
$r_B/10$) as function of the mean baryon density $n_b$.}
\end{figure}
\begin{figure}
\includegraphics[width=3.5in]{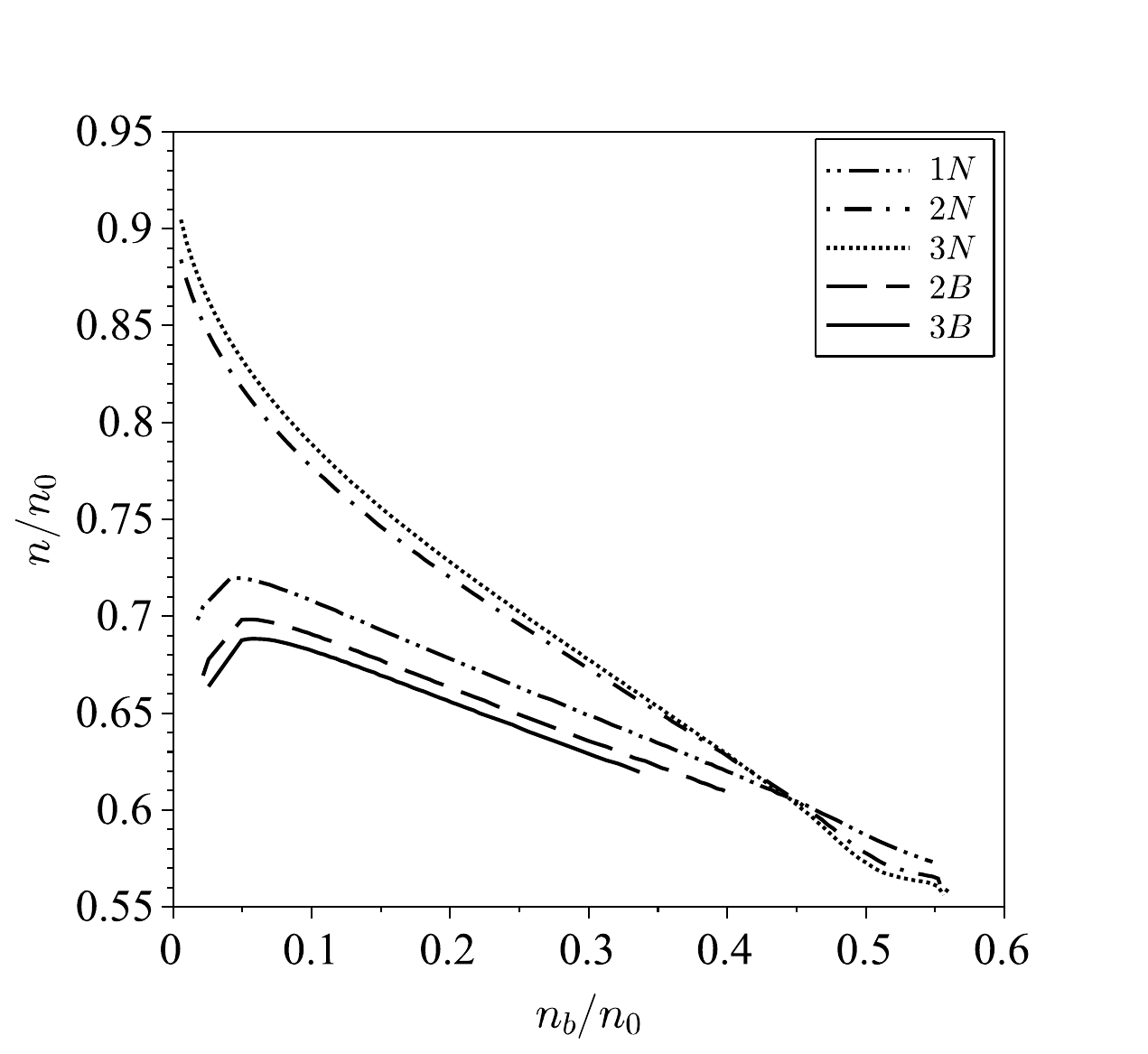}
\caption{Solution to the basic equations of equilibrium, Eqs. (\ref{var_1})-(\ref{var_4}): The baryon number density inside the nucleus and outside
 the bubble as function of the mean baryon density $n_b$.}
\end{figure}
\begin{figure}
\includegraphics[width=3.5in]{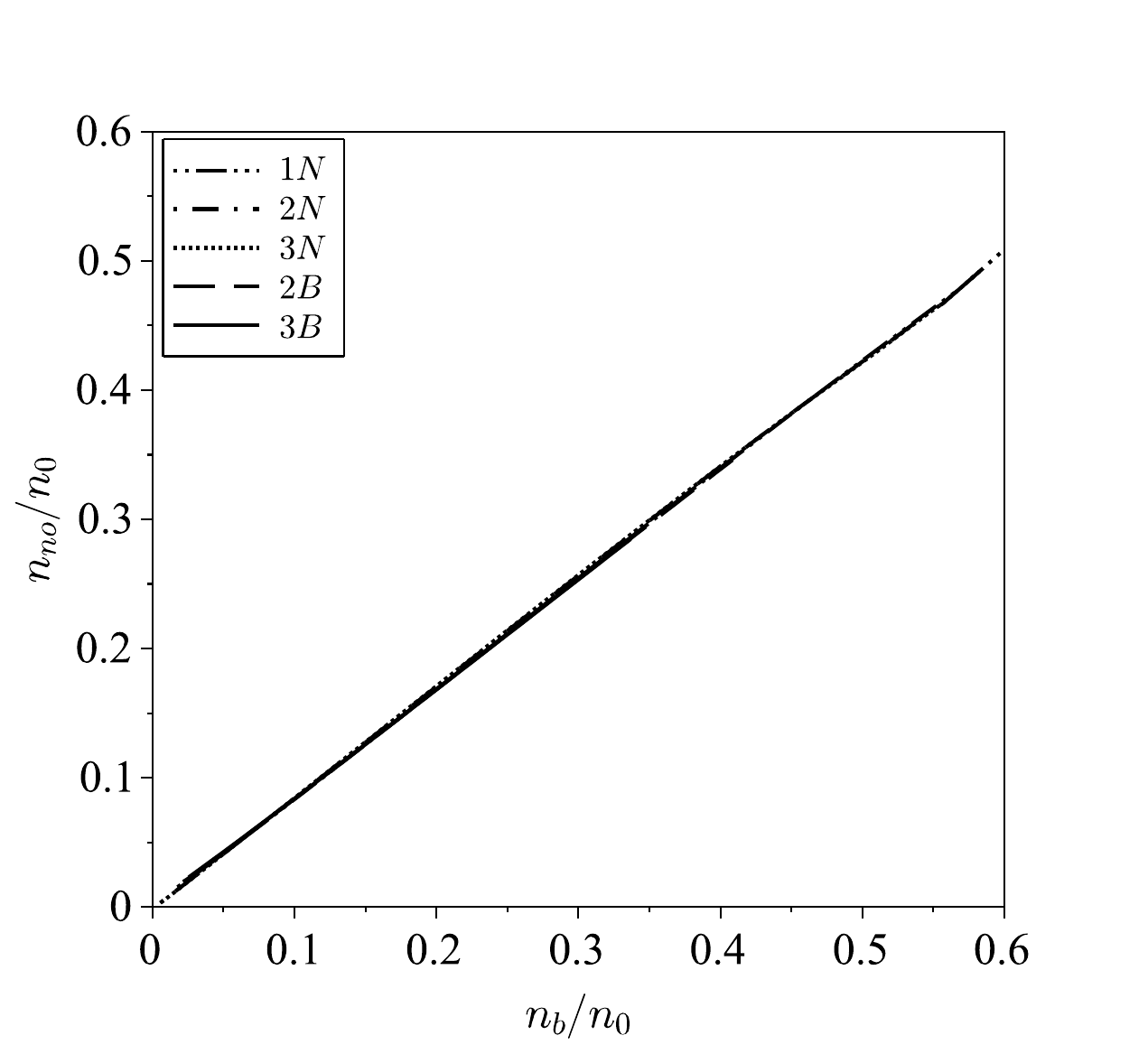}
\caption{Solution to the basic equations of equilibrium, Eqs. (\ref{var_1})-(\ref{var_4}): The baryon number density $n_{no}$ of pure neutron
matter outside the nucleus and inside the bubble as function of the mean baryon density $n_b$.}
\end{figure}

\subsection{Proton workfunction}
The variables shown in Figs. 1--4 represent a complete and unique solution to the basic equations, however, the appearance of pasta structures is
not explicit.
In order to reveal the appearance of pasta phases, as a first step, we compute the difference between the proton chemical potential inside and outside of the nuclear cluster (Eqs. (\ref{Mupi}), (\ref{Mupi_BUB}) and (\ref{Mupo})),
\begin{equation}\label{def_deltaMuP}
      \Delta\mu_p=\mu_{po}-\mu_{pi},
\end{equation}
which represents the workfunction of protons, which is a potential energy barrier created by the cluster, that prevents protons from flowing out of the nuclear cluster into the pure neutron matter.

Figure 5 shows $\Delta\mu_p$ as function of the mean baryon density $n_b$, for five possible pasta phases in the inner crust.
\begin{figure}
\includegraphics[width=3.5in]{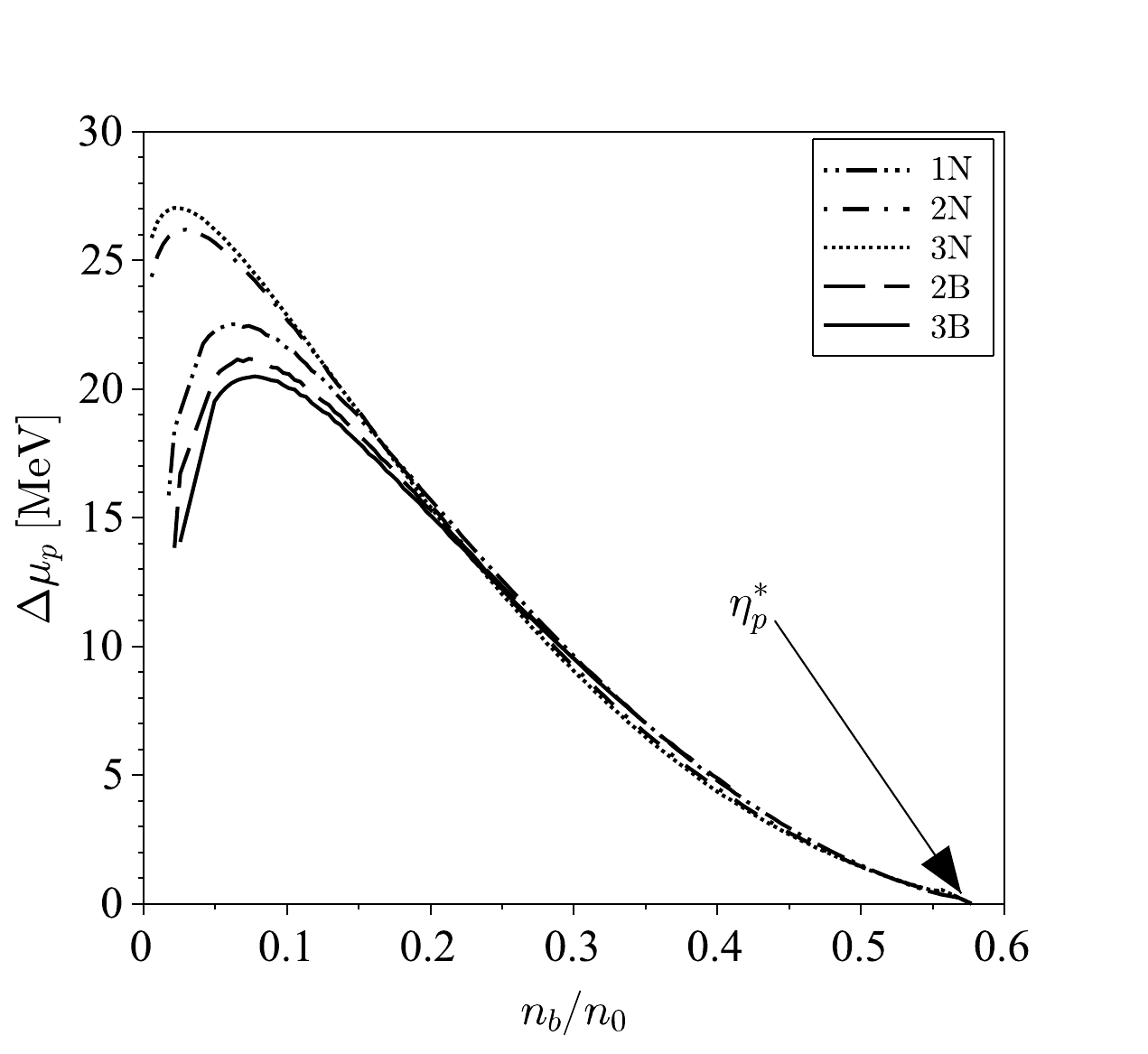}
\caption{The workfunction to transfer a proton from the nucleus to the surrounding pure neutron matter, Eq. (\ref{def_deltaMuP}), as function of
the mean baryon density $n_b$, calculated from the results shown in Figs. 1-4. The arrow marks the mean baryon density $\eta_p^{*}n_0$ corresponding
 to the proton drip.}
\end{figure}
From Fig. 5, we observe that at low $n_b$, the barrier is strong enough to ensure a complete insulation of the pure neutron matter from the proton  currents; this fact was also noticed by Zhang and Pethick in \cite{ZhangPethick2021} who assumed $\Delta\mu_p\sim 6$ MeV.
We find, however, that at densities $n_b$ higher than $\sim0.4n_0$ the barrier smoothly becomes less than 5 MeV, indicating that in some regions of pasta phases, the proton tunneling smoothly becomes important and in this case the pure neutron liquid does not provide efficient insulation against the tunneling proton supercurrents.

Ultimately, at $n_b/n_0=\eta_p^{*}\sim0.58$ the barrier vanishes, which implies that for $n_b/n_0\geq\eta_p^{*}$ the protons can freely leave the nucleus and enter the pure neutron liquid.
Our liquid drop model calculations using the Sk$\chi450$ interaction predict that the proton drip occurs at a mean baryon density $\eta_p^{*}n_0$, which is larger than the critical density $\eta_{\rm uni}^{*}n_0$ (above which the ground state is the uniform nuclear matter).
For this interaction, the value of the critical density is $n_b/n_0=\eta_{\rm uni}^{*}=0.5508$, which is computed using Eq. (30) and Table V of \cite{LimHolt2017} and reported in Table \ref{table3} of the present work.
This prediction of our model,
\begin{equation}\label{etastaretauni}
  \eta_{\rm uni}^{*}<\eta_{\rm p}^{*},
\end{equation}
has been obtained assuming that the second phase is the pure neutron liquid.
With this assumption, the surface energy is a function of the proton fraction only inside the nucleus (see Eqs. (\ref{def_ws}) and (\ref{def_ws})).
However, in the situation, not considered in this work, when the second phase is a dilute nuclear matter \cite{Vinas1998,douchin00,PethickEtal1995,KellerEtal2024}, the present liquid drop model should be extended in order to take into account nonzero proton fractions both inside and outside the nucleus.

\section{A simple model of superconductivity}
\label{Sec_SuperconductivityLDM}
In observable neutron stars, superconductivity of the pasta phases is important in the context of interaction between the inner crust and the stellar magnetic induction.
A proper description should be capable to account for the smeared character of profiles of the nucleon densities and for other microscopic features of the pairing problem.
Thus, the thermodynamic properties such as the critical magnetic field of the pasta phases, the proton-proton correlation length and the pairing energy, can be studied quantitatively.
While this topic will be addressed in the future work, here, we introduce a simple model, which is expected to qualitatively grasp the basic physical properties without a need to develop a complicated theoretical framework.

In our model, the proton density profile is sharp (it is zero outside and finite and uniform inside the nuclear cluster).
Similarly, the superconducting density is assumed to have a sharp profile.
We evaluate the coherence length and the London penetration depth associated with matter inside the nuclear cluster using the results obtained in pairing calculations in uniform nuclear matter.
The input data for our model is reported in Fig. 7 of \cite{LimHolt2021} for the proton pairing gap energy in beta equilibrium.
Thus, energy gap has been obtained from the pairing calculations with the same underlying nuclear interaction, as has been used for the structure calculations.

The pairing energy contribution to the total energy is calculated from thermodynamic arguments.
The excess (negative) energy associated with pairing is proportional to the gap energy and the number of paired particles at given temperature.
This leads to the following expression for the pairing energy density in liquid drop model averaged over the Wigner-Seitz cell at low temperature:
\begin{eqnarray}
  \label{def_w_pairing} && w_{p} = - \kappa u n x \Delta_p, \quad (\rm nucleus\; phases) \\
  \label{def_w_pairing_BUB} && w_{p}^{\rm bub} = - \kappa (1-u^{\rm bub}) n x \Delta_p, \quad (\rm bubble\; phases)
\end{eqnarray}
where $\kappa$ is a number that depends on microscopic properties and $\Delta_p$ is the value of the gap energy inside the nuclear cluster.
The quantities appearing in Eqs. (\ref{def_w_pairing}) and (\ref{def_w_pairing_BUB}) are constant numbers and have no spatial dependence within the unit cell.
Notice that the physical prediction for the pairing energy density that follows from the expression used by Pizzochero, Viverit and Broglia in Eq. (3) of \cite{Pizzochero1997} and later on by Pearson and Chamel in Eq. (14) of \cite{PearsonChamel2022}, is equivalent to the prediction following from our Eq. (\ref{def_w_pairing}) with a certain choice for $\kappa$.
This becomes evident as soon as Eq. (3) of \cite{Pizzochero1997} is spatially averaged over the Wigner-Seitz cell.
It is easy to see that in this case
\begin{equation}\label{kappa_pairing}
  \kappa = \frac{3}{8}\frac{\Delta_p}{E_{Fp}},
\end{equation}
where $E_{Fp}=\hbar^2(3\pi^2nx)^{2/3}/2{\tilde m}_p$ with ${\tilde m}_p$ given in Eq. (\ref{eqA12}).

\section{Analysis of numerical results}
\label{Sec_AnalysisNumericalResults}
\subsection{Pasta configurations with lowest internal energy and polymorphism}
Here, we focus on two questions. (i) What is the pasta phase with the minimum internal energy? (ii) Which pasta phases have energy per baryon that is larger than the minimum by less than the thermal energy?

Inserting the solutions shown in Figs. 1--4 into the expressions for the energy density, Eqs. (\ref{def_wtot}) and (\ref{def_wtot_BUB}), we
calculate the total internal energy density for each of the five pasta phases.
Next, we determine the ground-state configuration at zero temperature, which is the one with the lowest internal energy density at fixed total baryon number and fixed temperature.
Notice that for fixed $n_b$, as in our approach, the picture in terms of the energy per baryon is equivalent, up to a constant scaling factor, to the picture in terms of the energy density.
As a next step, we find all the possible structures whose energy per baryon differs from the ground state by less than the thermal energy $k_BT$.

It is worth noting that to find the most thermodynamically favorable state of the system, one needs to minimize the thermodynamic potential $\Phi$.
For this purpose, it is necessary to define a set of thermodynamic variables, which imposes restrictions on the form of $\Phi$.
For instance, in this work the thermodynamic variables are the total number of baryons $N_b$ in the unit cell with volume $V_c$ and temperature $T$.
In this case, $\Phi(N_b,T)=E_{\rm tot}(N_b)-k_BTS(N_b,T)V_c$, where $E_{\rm tot}(N_b)$ is the internal energy per unit cell and $S(N_b,T)$ is the total entropy of the system per unit cell evaluated at given $N_b$ and $T$.
Alternatively, an equivalent description can be given in terms of the total pressure $P$ and temperature $T$ and in this case $\Phi(P,T)=E_{\rm tot}(P) -k_BTS(P,T)V_c + PV_c$, where the internal energy and entropy are calculated for given $P$ and $T$.
However, temperatures in the range of $10^8$--$10^9$ K are very small compared to the typical nuclear energy scale of 1 MeV, thus,
the relevant quantity is the internal energy at $T=0$.
In this work we focus solely on the internal energy, while leaving the thermal effects for the future work.

Our results are displayed in Fig. 6, where the minimum internal energy is shown by black solid line, while the dot markers show the solutions with energy per baryon that is larger than the minimum energy per baryon, by less than the thermal energy $k_BT$, for three different temperatures.
The vertical blue lines show the critical densities $\eta_p^{*}$ and $\eta_{\rm uni}^{*}$ defined in Eq. (\ref{def_eta_uni_star}) and in Fig. 5.
The calculations for $n_b/n_0>0.6$ have been done on a sparser array of values since the uniform matter is the only possible solution in this case.
\begin{figure}
\includegraphics[width=3.5in]{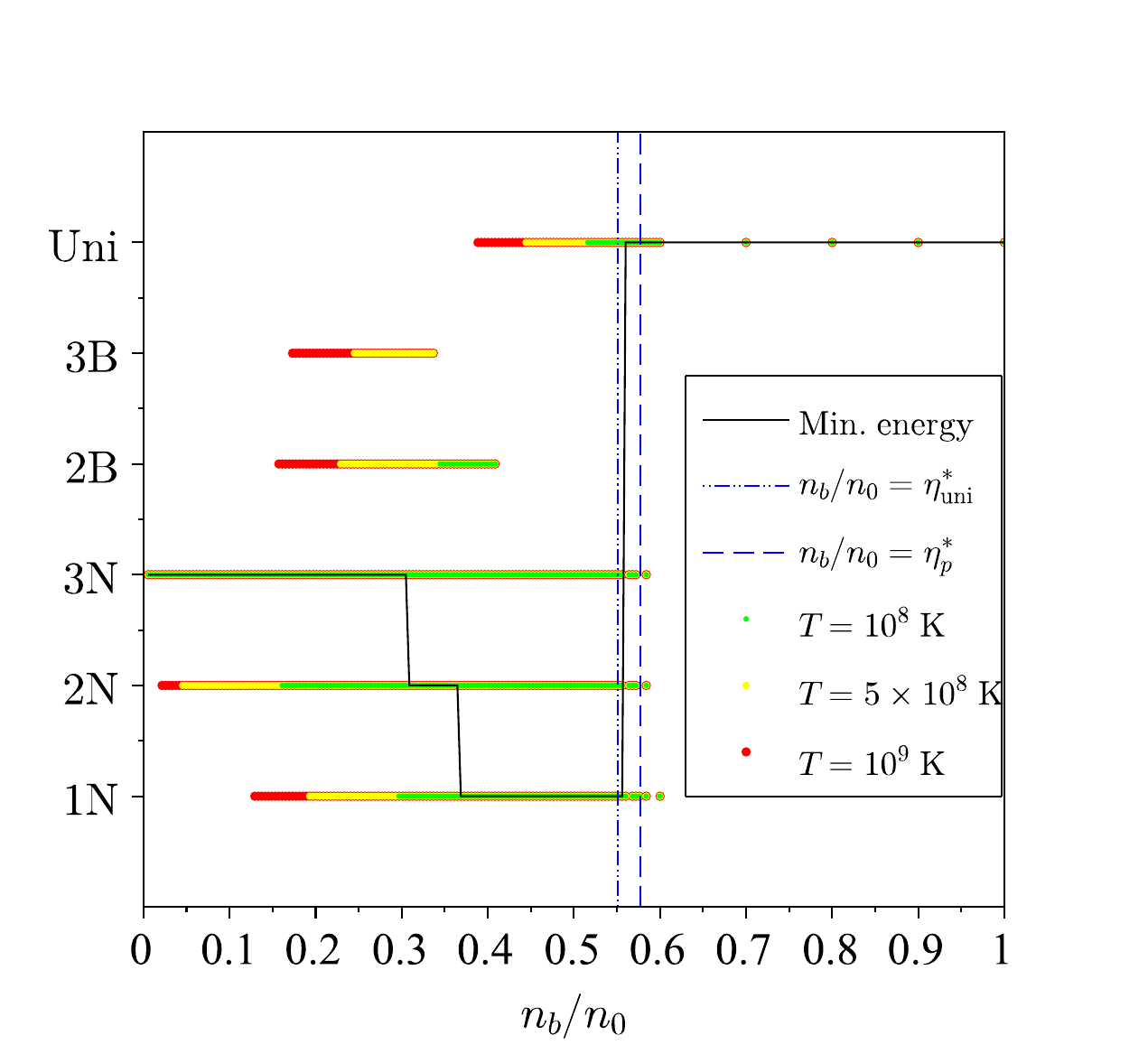}
\caption{(Color online) Thermodynamically allowed pasta configurations at three different temperatures $T$ taking on the values $10^9$ K (red),
$5\times10^8$ K (gold) or $10^8$ K (green).
Black solid line shows the the ground-state characterized by the minimum internal energy per baryon calculated from Eqs. (\ref{wtot}) and (\ref{wtot_BUB}).
The transitions between various pasta phases are seen as jumps of the black solid line, where the almost vertical pieces represent the absence of
the data because of the finite step size of the numerical grid for $n_b$.
The color markers indicate the excited configurations, which satisfy Eqs. (\ref{var_1})-(\ref{var_4}), while their internal energy per baryon is larger than the minimum internal energy by less than $k_BT$.
The dash-dotted line marks the critical value of $n_b$ at which uniform nuclear matter becomes unstable with respect to long wavelength
hydrodynamic fluctuations (see Eq. (\ref{def_eta_uni_star}) for details).
The dashed line marks the critical value of $n_b$ at which protons start dripping out from the nuclei (see Fig. 5 for details).}
\end{figure}

Figure 6 suggests that even in the low-temperature regime at $T\sim10^8$ K, the state of a real matter is likely a polymorphic structure, in which locally, nuclear clusters with various types of clusters (either of 1N, 2N, 3N, 2B and 3B or uniform) coexist.
It is known from the material science that the process of crystallization might lead to simultaneous formation of various crystal structures with close value of the thermodynamic potential, ending up with a polymorphic rather than a pure crystal structure.
A similar scenario for the crust was also anticipated in \cite{NewtonEtal2022}.

For instance, consider the case when $T=10^8$ K.
As it follows from our calculations, for $n_b$ below $\sim0.16n_0$, the system energetically favors only 3N phase.
However, for $n_b$ above $\sim0.16n_0$ and below $\sim0.297n_0$, the system favors both 3N and 2N phases.
For $n_b$ above $\sim0.297n_0$ and below $\sim0.345n_0$, or above $\sim0.41n_0$ and below $\sim0.515n_0$, the system favors 3N, 2N and 1N phases.
For $n_b$ above $\sim0.345n_0$ and below $\sim0.409n_0$, the favored phases are 3N, 2N, 1N and 2B phases.
Interestingly, for $n_b$ above $\sim0.515n_0$ and below $n_0\eta_{\rm uni}^{*}$, the system favors 3N, 2N, 1N and the uniform phases.

\subsection{Characteristic physical lengths of protons}
We start by noticing that Fig. 6 shows that lasagna is the ground state in a significant portion of the inner crust.
In the weak pairing Bardeen-Cooper-Schrieffer approximation, the coherence length, which can be understood as the spatial size of the Cooper
pair, reads
\begin{equation}\label{def_xi}
\xi(0)=\frac{\hbar v_{Fp}}{2\gamma\Delta_p},
\end{equation}
where $(0)$ implies that we neglect the thermal exponential suppression of $\xi$ by virtue of the condition
\begin{equation}\label{Tcondition}
\Delta_p>k_BT\sim0.01\,{\rm MeV}.
\end{equation}
In Eq. (\ref{def_xi}), the proton Fermi velocity is $v_{Fp}=\hbar k_p/m_p$, the Euler constant is $\ln\gamma=0.577$ and $\Delta_p=\Delta_p(n,x)$
is the superconducting energy gap.
The proton Fermi wavenumber reads
\begin{equation}\label{def_kFp}
  k_{Fp}=(3\pi^2 Y_pn_b)^{1/3},
\end{equation}
where the quantity $Y_p$ (the proton fraction averaged over the Wigner-Seitz cell) utilized in \cite{LimHolt2021} is given in variables of the present paper as
\begin{equation}\label{YpX}
    Y_p=unx/n_b.
\end{equation}

In order to reveal what part of lasagna is expected to be superconducting, we plot in Fig. 7 the superconducting gap energy as function of
the mean baryon density across the entire density range of lasagna.
\begin{figure}
\includegraphics[width=3.5in]{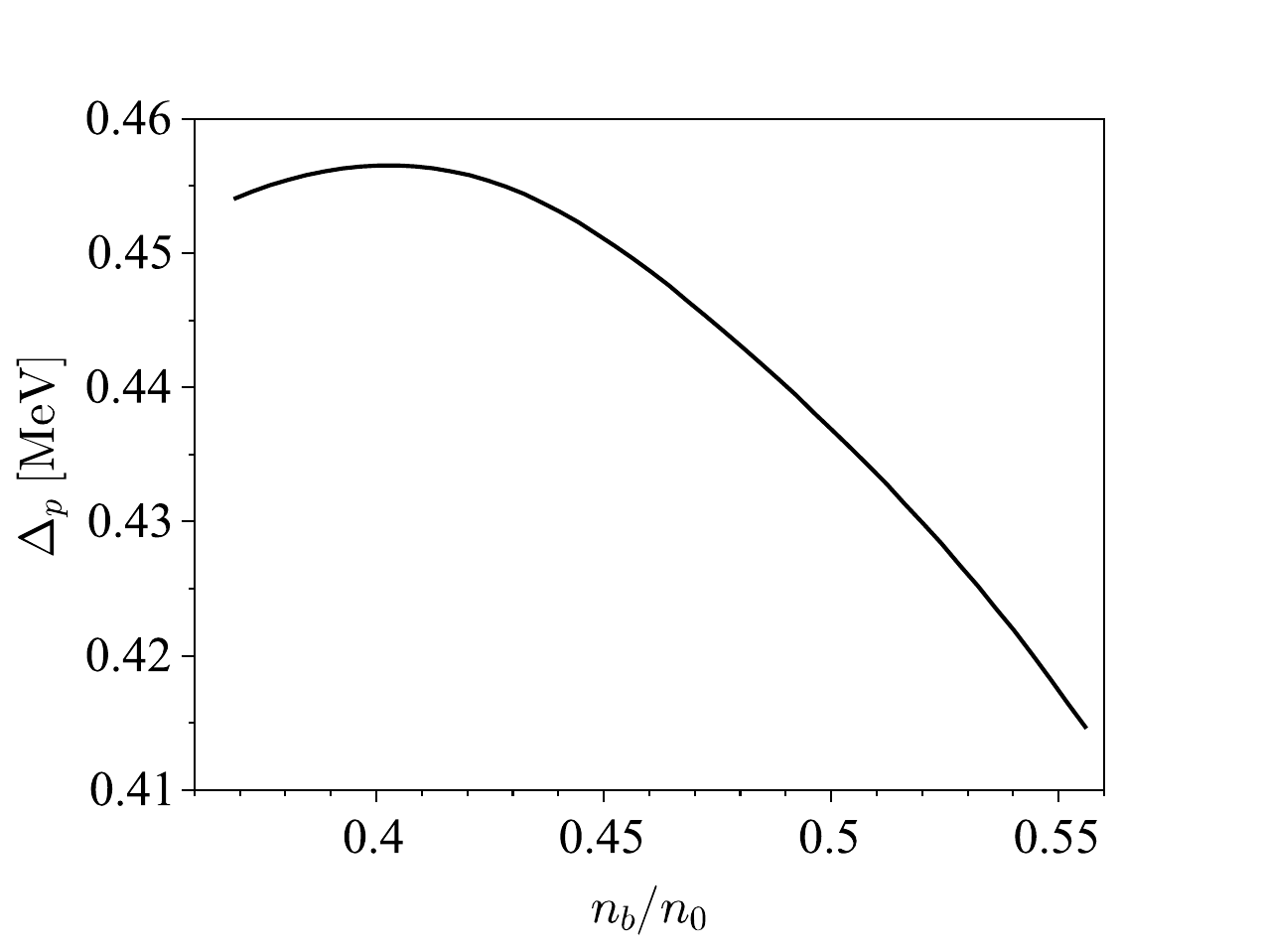}
\caption{The superconducting energy gap $\Delta_p$ in lasagna, adopted from Fig. 7 of \cite{LimHolt2021}, where the $k_{Fp}$-dependence has been converted into the $n_b$-dependence with the help of the functions shown in Figs. 1--4 and Eq. (\ref{def_kFp}).}
\end{figure}
Figure 7 shows that at typical $k_BT\sim0.01$ MeV, the thermodynamic condition given in Eq. (\ref{Tcondition}) does hold in the entire domain of lasagna.

The basic magnetic properties of superconducting matter in lasagna can be characterized by the ratio $d_L/\xi$ between the interlayer spacing $d_L$, where
\begin{equation}\label{def_dL}
d_L=2(r_c-r_N)=2(u^{-1}-1)r_N,\quad({\rm 1D\;nuclei})
\end{equation}
and the coherence length of protons $\xi$, Eq. (\ref{def_xi}), and by the Ginzburg-Landau (GL) parameter $\lambda/\xi$, where
\begin{equation}\label{def_lambdaL}
\lambda^2(0)=\frac{m_pc^2}{4\pi e^2 n_{p}(0)}
\end{equation}
defines the square of the London penetration depth of the magnetic field for the bulk matter.
By using in Eq. (\ref{def_lambdaL}) the value of the superconducting proton density averaged over the Wigner-Seitz cell, we obtain the lower bound on the magnetic penetration depth.
If there are neutrons between the slabs, the magnetic penetration depth might increase because in this case the superconductor becomes topologically different from uniform superconductor, as show panels (a) and (b) in Fig. 8.
\begin{figure}
\includegraphics[width=3.5in]{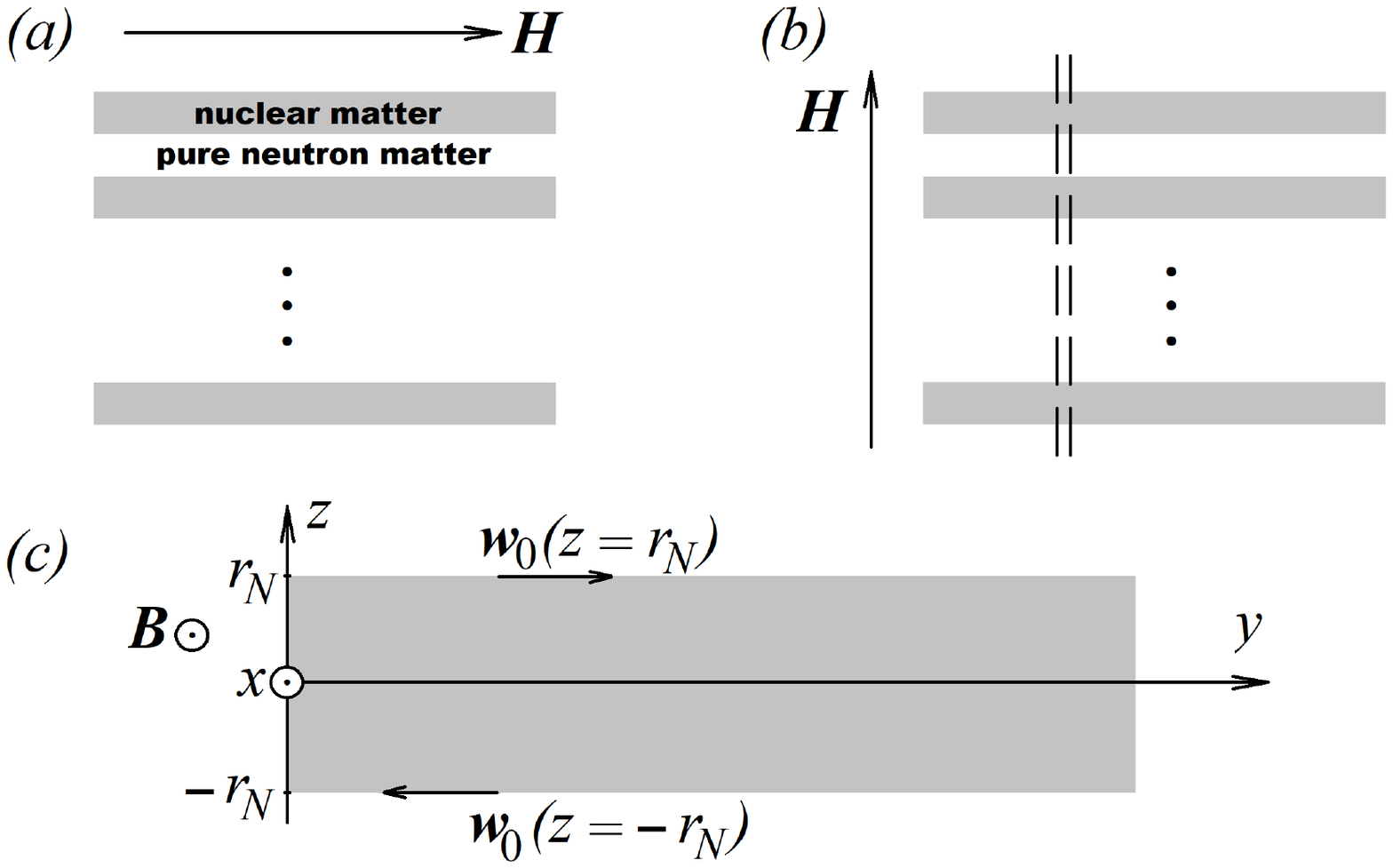}
\caption{Schematic representation of a unit length of ideally ordered lasagna, which currently is not excluded as a real state of the inner
crust.
Three dots in panels (a) and (b) express the presence of the ideal lattice of slabs.
Panel (a) shows a configuration with the magnetic field parallel to the slabs.
Panel (b) displays a configuration with the magnetic field parallel to the slabs with dashed lines representing the normal core of the magnetic
flux tube in the proton superconductor.
Panel (c) depicts a single slab within the magnetic induction with arrows representing the neutron-proton momentum lag vector at equilibrium.}
\end{figure}

The magnetic flux through lasagna is nonquantized in the case shown in Fig. 8 (a) and is quantized in Fig. 8 (b).
Notice that the anisotropy of lasagna induces a direction-dependence of the London penetration depth, in contrast to the isotropic
Eq. (\ref{def_lambdaL}).

Figure 9 shows the dependence of the parameters $d_L$, $r_N$ and $\xi$ in lasagna on the mean baryon density $n_b$, where we also plot the
scaled London penetration depth $\lambda$, the characteristic screening length $k_{TFe}^{-1}$ in the electron background and the inverse Fermi wavenumber of the electrons $k_{e}^{-1}$.
Here,
\begin{equation}\label{def_kTFe}
k_{TFe}^{2}=4\pi e^2\frac{\partial n_e}{\partial \mu_e}.
\end{equation}
with $n_e$ and $\mu_e$ given in Eqs. (\ref{def_ne}) and (\ref{def_mu_cluster}), correspondingly, so ${\partial n_e}/{\partial \mu_e}=k_e^2/\pi^2\hbar c$, where
\begin{equation}\label{def_ke}
k_e^3=3\pi^2 n_e.
\end{equation}
In Fig. 9, we observe that $\lambda/\xi>\sqrt{2}$ in the entire range of $n_b$ corresponding to lasagna and therefore lasagna is expected to be in the type-II regime of superconductivity.
\begin{figure}
\includegraphics[width=3.5in]{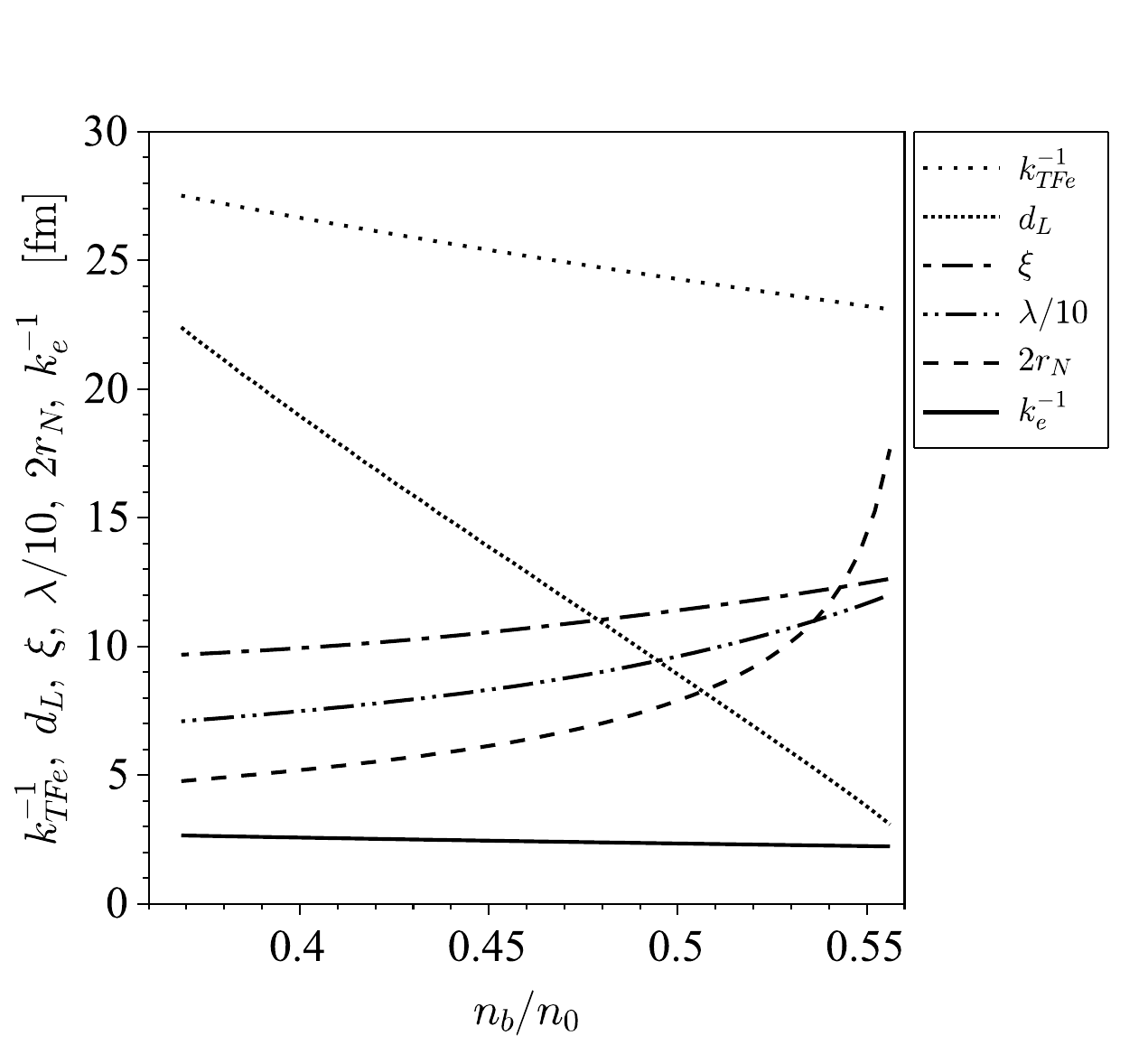}
\caption{The characteristic parameters of proton superconductor in lasagna calculated from Figs. 1-4 and supplemented by the data on
$\Delta_p(k_{Fp})$ from Fig. 7 of \cite{LimHolt2021}.
The distance between the nearest interfaces of two adjacent slabs is $d_L$, $\lambda$ is the lower bound on the London penetration depth of the magnetic field into the superconducting protons, $\xi$ is the coherence length of protons at temperatures well below $\Delta_p/k_B$, $2r_N$ is the width of the slab (also given in Fig. 2), $k_{TFe}^{-1}$ is the characteristic screening length in the electron background and $k_{e}^{-1}$ is the electron Fermi wavenumber.}
\end{figure}

It is seen in Fig. 9, that the individual slabs have 2-dimensional character for $0.37n_0\lesssim n_b\lesssim0.54n_0$ (when $\xi>2r_N)$, while for $0.54n_0\lesssim n_b\lesssim 0.56n_0$ the slabs are effectively 3-dimensional (when $\xi<2r_N$).
The information about dimensional character of the slabs is essential for calculations involving the magnetic vortex structure and the critical magnetic field.

Another aspect of the problem is the interaction between the neighboring slabs, which can be quantified by the ratio of $\xi$ to $d_L$.
In the range of $0.37n_0\lesssim n_b\lesssim0.48n_0$ (when $\xi<d_L$), the slabs are not significantly coupled by the Josephson effect and therefore, the appropriate picture is the layered discreet superconductivity.
However, at higher densities, in the range of $0.48n_0\lesssim n_b\lesssim0.56n_0$ (when $\xi>d_L$), the appropriate picture is the anisotropic 3-dimensional superconductivity.

In a recent work by Zhang and Pethick \cite{ZhangPethick2021}, it was assumed that the individual slabs have 2-dimensional character and that the neighboring slabs are coupled by the Josephson effect and, thus, the 3-dimensional anisotropic picture of superconductivity was used.
Figure 9 shows that the approach used in \cite{ZhangPethick2021} is validated by our numerical results in the range of $0.48n_0\lesssim n_b\lesssim0.54n_0$.
In contrast, in an earlier work by Kobyakov \cite{Kobyakov2018}, it was assumed that the neighboring slabs are not coupled by the Josephson effect and thus, the discreet layered picture of superconductivity was used.
This approach is validated in the range of $0.37n_0\lesssim n_b\lesssim0.48n_0$.

\section{Astrophysical implications}
\label{Sec_Astrophysical}
\subsection{Overview of calculation of stress in superconducting matter of neutron stars}
Tensor of the momentum and energy, or briefly, the stress tensor, contains information about hydrodynamic and magnetic content in the system.
It is useful for studies of pressure forces and of interaction of the fluid with the magnetic field.
Often, it is convenient to study the stress tensor for some specific form of motion and specific initial conditions for the magnetic field.
For instance, the case that can be loosely described as magnetostatic-hydrostatic was studied by Easson and Pethick in \cite{EassonPethick1977}, while the magnetostatic-hydrodynamic situation was studied by Sedrakian and Sedrakian in \cite{Sedrakian1995} in the context of the uniform nuclear matter in the core of neutron stars and later by Kobyakov in \cite{Kobyakov2018} in the context of nonuniform matter in the inner crust of neutron stars.

The pioneering work \cite{EassonPethick1977} was done for the case of a uniform isotropic superconductor threaded by the quantized magnetic flux lines, while there is no macroscopic flow of matter.
In the following work, scenarios involving some form of a macroscopic fluid flow were studied.
The approach of Sedrakian and Sedrakian \cite{Sedrakian1995} focused on interaction of the superconducting protons with the superfluid rotation of neutrons, which generates the magnetization due to the entrainment effect.
Intrinsic momentum coupling between superfluid neutrons and superconducting protons is a phenomenon known as the \emph{entrainment} in the
superconducting-superfluid uniform nuclear matter.
The entrainment contribution to the total energy density depends on the electromagnetic vector potential $\mathbf{A}$ through the gauge-invariant difference of $\mathbf{A}$ and the gradient of the phase of the proton superconducting order parameter $\nabla\phi_p$.
As a next step, Kobyakov \cite{Kobyakov2018} studied some effects of a macroscopic flow of superfluid neutrons in lasagna.
Notice that the magnetic induction is almost equal to the magnetic field inside the slab, because, as shows Fig. 9, the condition
\begin{equation}\label{Llambda_rN}
  \frac{\lambda}{2r_N}\gg1
\end{equation}
holds for the entire domain of lasagna.

The mutual rotational evolution of the nonsuperfluid and superfluid particles naturally includes situations when there is an adiabatic perturbation of the macroscopic flow of neutrons in the rest frame of reference of the center-of-mass of the slab.
A relevant question in this case is, whether the perturbation of the neutron superflow in the magnetic field $\mathbf{H}$ would lead to a magnetohydrodynamic stress?

This question has been raised in the earlier work \cite{Kobyakov2018}.
The basic idea in \cite{Kobyakov2018} is that the system is a toroidal volume containing ideal lasagna.
The slabs are oriented along the field lines of the toroidal component $\mathbf{B}$ of the stellar magnetic induction, as schematically shown in Fig. 10 (a).
\begin{figure}
\includegraphics[width=3.5in]{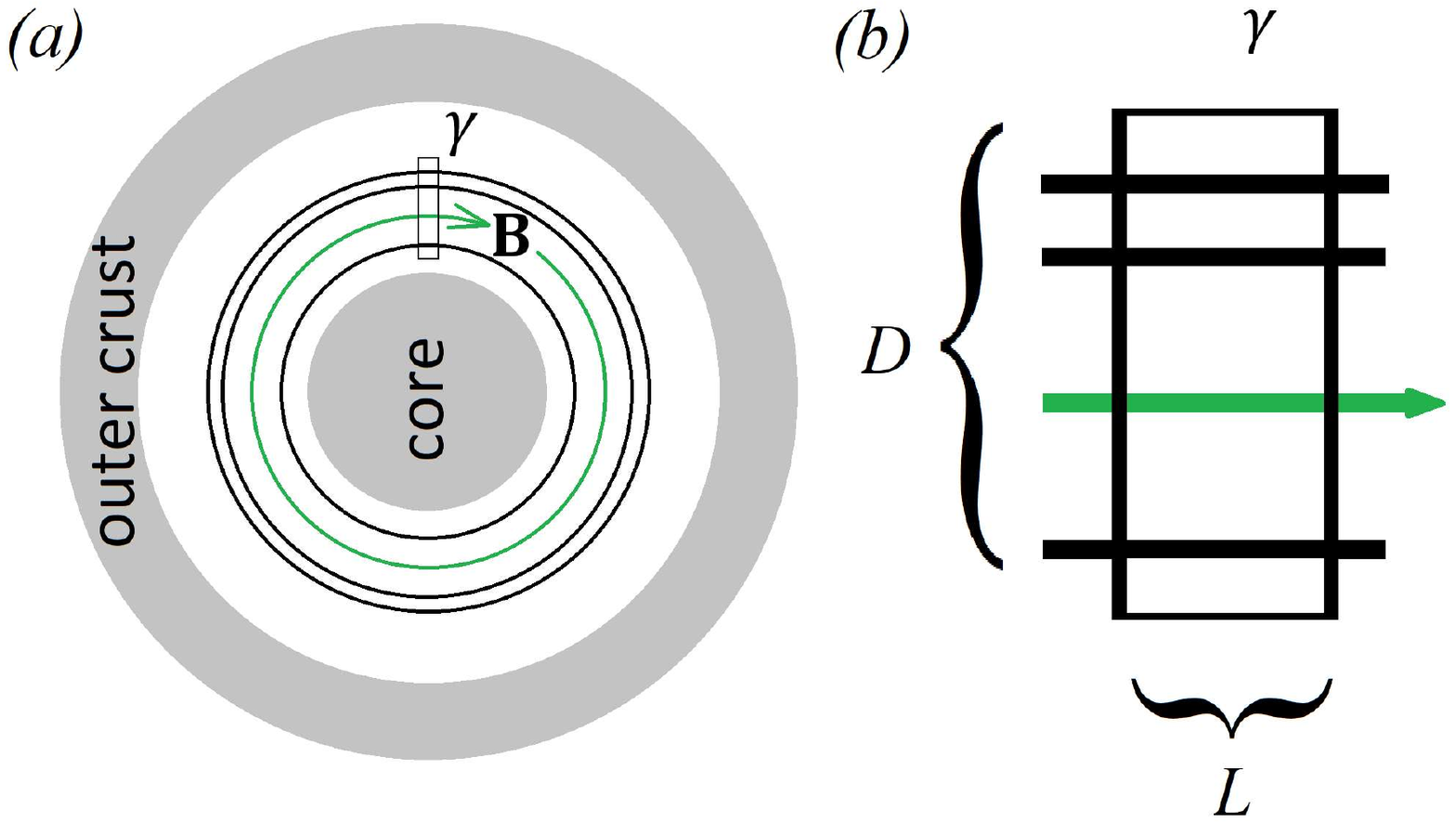}
\caption{(Color online). Panel (a): A schematic representation (not to the scale) of the magnetic-equatorial cross-section of the star, highlighting lasagna located between the outer crust (when mean baryon density $n_b$ is smaller than the neutron drip density) and the core (when $n_b>n_0\eta_{\rm uni}^{*}$) and threaded by the toroidal lines of the stellar magnetic induction $\mathbf{B}$.
Thin solid rectangle denoted as $\gamma$ represents the cut piece of the toroidal volume shown in panel (a) of Fig. 8.
Panel (b): Enlarged view of 2-dimensional projection of 3-dimensional piece of volume $\gamma$ ``cut'' from panel (a), where $D$ is the height of the column discussed in Eq. (\ref{Ftot1d_nxu}).
The planar nature of the system seen in panel (b) and in Fig. 8 is due to the very large curvature radius of the structure as compared with the structure lattice period.
The periodic boundary conditions in the plane parallel to the slabs, which are used in our calculations are essential for existence of the neutron superflow.
The parameter $L$ was introduced in Eq. (\ref{def_VN}).}
\end{figure}

Let us now ``cut'' a piece of lasagna, with spatial dimensions $D\times L^2$, as schematically shows thin rectangle in Fig. 10 (a).
This rectangle is enlarged in Fig. 10 (b) in order to clearly define the physical system under consideration.
By cutting a piece with $2r_N\ll L\ll R_{*}$, where $R_{*}$ is the stellar radius, we convert the physical problem of the inner crust shown in Fig. 10 (a) into an idealized model where the slabs are strictly flat as shown in Fig. 10 (b).
This situation corresponds to the model shown in Fig. 8 (a).
We assume periodic boundary conditions for the nucleon number currents in the plane parallel to the slab, so that the neutron and proton flows parallel to the slab surface are possible.
If the condition $d_L>\xi$ is satisfied, the slabs in Fig. 8 (a) are independent.
We assume this condition holds and therefore it is sufficient to consider only a single slab, as shown in Fig. 8 (c).

Physically, a single slab in the magnetic field develops electric currents flowing inside the superconducting domain parallel to the surface and attempting to screen the magnetic field.
Because of the form of the equilibrium vector potential $\mathbf{A}_0$, the electric currents at the opposite surfaces of the slab have opposite directions.
The electrons can be assumed static, because the opposite surfaces of the slab are separated by less than the screening length $k_{TFe}^{-1}$ shown in Fig. 9.

Following \cite{Kobyakov2018}, we assume that the temperature is low enough, so that the superfluid nucleon densities are equal to the bulk nucleon densities $n_\alpha$, $(\alpha=p,n)$.
For convenience, let us remind that the superconducting and superfluid S-wave order parameters, which can be generally written as $\sqrt{n_\alpha}\exp{\mathrm{i}\phi_\alpha}$, consist of the superfluid density $n_\alpha$ and the superfluid phase $\phi_\alpha$.
The fluid specific momenta are $\mathbf{p}_\alpha=\hbar\nabla\phi_\alpha$.
The lag of superfluid momenta scaled by the particle mass, is defined as
\begin{equation}\label{def_w}
  \mathbf{w}=\frac{1}{m}\left[ \left(\mathbf{p}_p-\frac{e}{c}\mathbf{A} \right) - \mathbf{p}_n\right].
\end{equation}
The energy density of the system shown in Fig. 8 (c) reads
\begin{equation}\label{def_Hmatttot}
  w_{\rm tot}^{\rm matt} = w^{\rm nuc} + w_{\rm kin}^{\rm nuc} + w_{\rm Coul}^{\rm ep} + w^{\rm e} + w^{\rm no},
\end{equation}
where $w^{\rm nuc}$ is given in Eq. (\ref{def_wnuc}), $w^{\rm e}$ is given in Eq. (\ref{def_we}), $w_{\rm Coul}^{\rm ep}=e(n_p-n_e)\Phi$, $\Phi$ is the electrostatic potential, and the density of kinetic energy of the nucleons is
\begin{equation}\label{def_wnuckin}
  w_{\rm kin}^{\rm nuc} = \frac{n_p{\pi}_p^2}{2m} + \frac{n_n{p}_n^2}{2m} - \frac{n_{np}}{2m}(\pi_p - \mathbf{p}_n)^2.
\end{equation}
Here, $\pmb{\pi}_{p}=\mathbf{p}_{p}-e\mathbf{A}/c$ is the gauge-invariant proton momentum, $\mathbf{A}$ is the electromagnetic vector potential and $m=m_p\approx m_n$ and $\mathbf{B}=\nabla\times\mathbf{A}$.

The equations of motion generated from the energy functional with density given in Eq. (\ref{def_Hmatttot}) have the form of the continuity equation and the Newtown's second law for each of the nucleon species \cite{Kobyakov2018}:
\begin{eqnarray}
\label{EqsConti}  \partial_t n_\alpha + \nabla\cdot\mathbf{J}_\alpha = 0, \\
\label{EqsMomen}  \partial_t \mathbf{p}_\alpha = -\nabla\frac{\partial w_{\rm tot}^{\rm matt}}{\partial n_\alpha},
\end{eqnarray}
where $\mathbf{J}_\alpha$ are the number currents of each species of nucleons.
Combining the momentum equations and using the incompressibility condition $\nabla\cdot\mathbf{J}_\alpha=0$, one arrives at the expression for the stress tensor.

Using the geometry shown in Fig. 8 (c), the equilibrium vector potential is $\mathbf{A}_0=-\mathbf{\hat{y}}B_0z$ and the unperturbed superfluid gauge-invariant momentum lag $\mathbf{w}_0$ reads
\begin{equation}\label{def_w0}
  \mathbf{w}_0=(e/mc)\mathbf{\hat{y}}B_0z,
\end{equation}
see equation (35) in \cite{Kobyakov2018}.
The total force between the magnetic field and the slab is zero in equilibrium.
This also can be seen from Eq. (\ref{Ftot2}) with $\mathbf{w}=\mathbf{w}_0$, or, $\delta\mathbf{w}=0$.

\subsection{Calculation of stress in a symmetric setting}
In the remainder of this section we will study perturbation of the force as a result of perturbation of the neutron superflow.
The expression for the force was given in Eq. (40) of \cite{Kobyakov2018} (see also a corrected expression in \cite{Kobyakov2023}).
As compared with the work done in \cite{Kobyakov2018}, here we will specify the quantitative evaluation of the force based on the microscopic parameters calculated in Sec. \ref{Sec_NumericalResultsLDM} and generalize the expression for the force for the case when $\mathbf{A}_0$ and perturbation $\delta\mathbf{w}$ are not parallel.
Following \cite{Kobyakov2018} and in the coordinates shown in Fig. 8 (c), the relevant force with $L=1$ cm reads
\begin{equation}\label{Ftot_fromEMtensor}
\delta^{(1)}\mathbf{F}_{\rm tot}^{1\,{\rm cm}^2\times2r_c}=\mathbf{\hat{z}}\int d^3\mathbf{r} \sum_k\nabla_k\Pi_{3k},
\end{equation}
where the stress tensor is
\begin{equation}
\label{def_Piik}\Pi_{ik}=J_{p\,k}\pi_{p\,i}+J_{n\,k}p_{n\,i}+\delta_{ik}P+\frac{1}{4\pi}(\delta_{ik}\frac{B^2}{2}-B_kB_i).
\end{equation}
The hydrodynamic pressure reads
\begin{equation}\label{def_P}
P=p-\rho_{np}^{*}\mathbf{w}^2/2,
\end{equation}
where $p$ is the pressure without the contributions due to the matter flows, as given in equation (32) in \cite{Kobyakov2018}.
$P$ is generally defined as the material contribution to $\Pi_{ik}$ associated with the isotropic tensor $\delta_{ik}$.
Here, $\rho_{np}^*=mn_{np}-mn_p\theta_p-mn_n\theta_n$ is the entrainment mass density (in the Skyrme model, $\theta_p=\theta_n=0$), $n_{np}$ is the entrainment number density and $P$ is the total physical pressure (the hydrodynamic pressure).
We evaluate the entrainment number density from (see Eq. (49) in \cite{KobyakovEtAl2017})
\begin{equation}\label{def_nnp}
  n_{np}=\frac{2}{n_0}\left(1-\frac{m}{m_0^{*}}\right)n_nn_p,
\end{equation}
where $n_p$ and $n_n$ are the nucleon number densities in uniform matter, $m_0^{*}$ is so-called nucleon \emph{isovector} effective mass, which reads (see Eq. (3.22) in \cite{ChabanatEtal1997})
\begin{equation}\label{def_m_eff}
m_0^{*}=\frac{m}{1+({m}/{4\hbar^2})n_0\left[t_1\left(x_1+2\right) + t_2\left(x_2+2\right)\right]}.
\end{equation}
The isovector effective mass is related to the effective mass given in Eq. (\ref{eqA12}) as following:
\begin{equation}\label{meff_relation}
  {m_0^{*}}=\left.\tilde{m_q}\right|_{n_q=0}.
\end{equation}

It remains to evaluate the total force as resultant of the forces applied to each of the slabs in the lattice (see Fig. 8 (a)).
From independence of the slabs, the total force acting on a 1 ${\rm cm}^2\times D$ column of lasagna is a sum of (integer part of) ${D}/{2r_c}$ equal terms, or
\begin{equation}\label{Ftot1d}
  \delta\mathbf{F}_{\rm tot}^{1\,{\rm cm}^2\times D}=\frac{D}{2r_c}\delta^{(1)}\mathbf{F}_{\rm tot}^{1\,{\rm cm}^2\times2r_c},
\end{equation}
where $D$ is the height of the column measured along the normal direction of the slabs, as shows panel (b) in Fig. 10.

Let us assume that the magnetic field $\mathbf{H}$ is pointed at an arbitrary angle $(\pi/2-\theta)$ relatively to the neutron velocity perturbation.
Let us remind that in \cite{Kobyakov2018}) the angle was $\pi/2$, or equivalently $\theta=0$.
The magnetic induction $\mathbf{B}$ is still restricted to the plane parallel to the slab surface and the momentum lag has the form
\begin{equation}\label{wperturb}
\mathbf{w}=\mathbf{w}_0+\delta\mathbf{w},
\end{equation}
where $\mathbf{w}_0$ is the equilibrium value and $\delta\mathbf{w}=\delta w(\mathbf{\hat{y}}\cos\theta+\mathbf{\hat{x}}\sin\theta)$ is the perturbation.
In order to find the force acting on a single slab (as shown in Fig. 8 (c)), we evaluate the perturbation of the stress tensor caused by the momentum lag perturbation in Eq. (\ref{Ftot_fromEMtensor}), and we obtain:
\begin{eqnarray}
\nonumber && \delta^{(1)}\mathbf{F}_{\rm tot}^{1\,{\rm cm}^2\times2r_c}=1\,{\rm cm}^2\times\left[P(z=r_N) - P(z=-r_N)\right]\mathbf{\hat{z}} \\
\label{Ftot2} && =-1\,{\rm cm}^2\times\mathbf{\hat{z}}\frac{e}{c}B_02r_Nn_{np0}\delta w\cos\theta,
\end{eqnarray}
where $n_{np0}$ is the entrainment number density inside the slab, $(\pi/2-\theta)$ is the angle between the magnetic field $\mathbf{B}_0$ and the vector of perturbation of the momentum lag $\delta\mathbf{w}$, which points at the same direction as the neutron velocity perturbation.

It is convenient to express the momentum lag in terms of the velocity lag $\delta v$, defined as
\begin{equation}\label{wandv}
\mathbf{v_p}-\mathbf{v_n} = (\mathbf{\hat{y}}\cos\theta+\mathbf{\hat{x}}\sin\theta)\delta v,
\end{equation}
via
\begin{equation}\label{dwdv}
\delta{w}=\frac{n_pn_{ni}}{n_{pp}n_{nn}-n_{np}^2}\delta v,
\end{equation}
with $n_{pp}=n_p-n_{np}$ and $n_{nn}=n_{ni}-n_{np}$.
In terms of the dimensionless variables used in this paper, we obtain
\begin{equation}\label{detnnp}
n_{pp}n_{nn}-n_{np}^2=n_0^2x(1-x)\eta^2\left[1-2\left(1-\frac{m}{m_0^{*}}\right)\eta\right],
\end{equation}
where $\eta$ is given in Eq. (\ref{def_eta}).
Combining Eqs. (\ref{Ftot2})-(\ref{detnnp}) and (\ref{Ftot1d}), we obtain
\begin{equation}\label{Ftot1d_nxu}
  \delta\mathbf{F}_{\rm tot}^{1\,{\rm cm}^2\times D}= -\mathbf{\hat{z}}\,2(1-\frac{m}{m_0^{*}})D\frac{e}{c}B_0n_0 u\frac{x\left(1-x\right)\eta^2}{1-2(1-\frac{m}{m_0^{*}})\eta}\delta v.
\end{equation}

We assume the magnetic field $B_0=5\times10^{14}$ G and the velocity lag between neutrons and protons $\delta v\sim1\;{\rm cm\,s}^{-1}$.
This value of $\delta v$ corresponds to a rotational lag between superfluid neutrons and the normal part of the star of the order of $10^{-6}$ s$^{-1}$.
In this case, Eq. (\ref{Ftot1d_nxu}) can be written as an engineering formula:
\begin{eqnarray}
\label{Ftot1d_nxu_engi} && \left|\delta\mathbf{F}_{\rm tot}^{1\,{\rm cm}^2\times D}\right| = 2.501\times10^{33}\;[{\rm dyn}] \\
\nonumber && \times\left|1-\frac{m}{m_0^{*}}\right| u\frac{x\left(1-x\right)\eta^2}{1-2(1-\frac{m}{m_0^{*}})\eta} \\
\nonumber && \times\left( \frac{D}{1\;{\rm cm}} \right) \left( \frac{B_0}{5\times 10^{14}\;{\rm G}} \right) \left( \frac{\delta v}{1\;{\rm cm\:s}^{-1}} \right).
\end{eqnarray}
Noticing that at given $n_b$, the variable $u$ is uniquely determined by the values of $n$ and $n_{no}$ given in Figs. 3 and 4, in Fig. 11 we plot the total force acting on a column of lasagna in the configuration shown in Figs. 8 (a), (c) for
$\theta=\pi/2$, at $D=1\;{\rm cm}$, $B_0=5\times10^{14}\;{\rm G}$, $\delta v=1\;{\rm cm\,s}^{-1}$, ${n_0}={0.1561\;{\rm fm}^{-3}}$ and $m_0^{*}/m=0.9466$.
\begin{figure}
\includegraphics[width=3.5in]{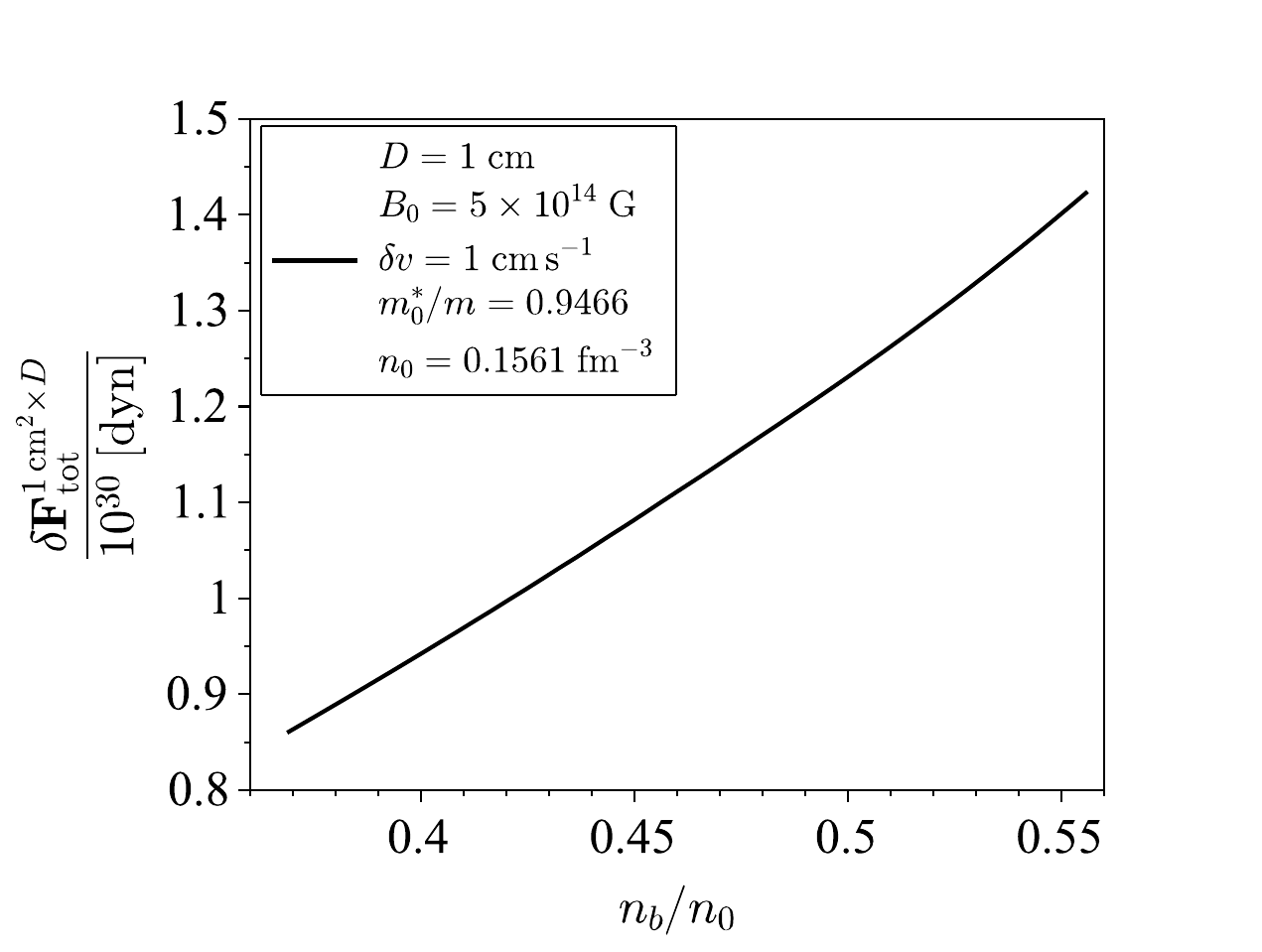}
\caption{Numerical evaluation of Eq. (\ref{Ftot1d_nxu_engi}) in lasagna with the variables $u$, $n$ and $x$ shown in Figs. 1-4 and with the other parameters specified in the inset.}
\end{figure}

The result presented in Fig. 11 corroborates the earlier conclusion regarding the possible magnitude of the force (see the corresponding discussion and Eq. (44) in \cite{Kobyakov2018}) and reveals a more detailed picture than existed earlier.

It is worth to emphasize that we have assumed $D\sim1\;{\rm cm}$ from the dimensional considerations and within the basic assumptions of the model, though likely $D\ll1$ cm in real neutron stars.
As to the velocity perturbation $\delta v\sim1\;{\rm cm\,s}^{-1}$, its magnitude has been assumed from the observational data on pulsars, which constrain the lag of the rotational frequencies of the superfluid and nonsuperfluid components to maximum values of the order of $10^{-6}\;{\rm s}^{-1}$ \cite{EysdenMelatos2010}.

\section{Conclusions}
We have solved the complete set of the variational equations of liquid drop model, Eqs. (\ref{var_1})--(\ref{var_4}).
We have computed self-consistently the surface tension of the interface that separates the nucleus from the pure neutron matter and the corresponding curvature correction, using the extended Thomas-Fermi approximation.
In our calculations, the pairing energy gap  has been extracted from
the data obtained on the basis of same underlying nuclear interaction, the effective chiral field theory,
as the one used in calculations of the bulk and the surface nuclear energies.

Our calculations show that inclusion of the curvature correction to the surface energy of nucleus changes the resulting energies of the pasta phases in such a way that the bubble phases completely disappear from the ground state, which is the state with the minimum internal energy per baryon among the entire set of pasta phases.
We identify a low-energy subset of pasta phases, which includes the minimum-energy phase (shown by black solid line in Fig. 6) and the excited low-energy phases (shown by three types of color markers in Fig 6, corresponding to $T$ equal to $10^8$, $5\times10^8$ and $10^9$ K).
The excited phases have energy per baryon not exceeding that of the minimum-energy phase by more than $k_BT$.

We interpret the low-energy set as phases that appear during the crystallization process owing to the thermodynamic fluctuations.
Thus, we conclude that the inner crust is likely to be polymorphic rather than a pure crystal.
The spatial and temporal extension of the excited phases is to be calculated in the future.
Appearance of the minimum-energy phases in the crust corresponds to the following pattern: 3N at lower mean baryon density $n_b$ followed by 2N, 1N and uniform at higher $n_b$.
Lasagna is found in a significant part of the crust.

We have introduced a simple model of superconductivity in lasagna incorporated into liquid drop model.
It follows from our numerical estimates that lasagna hosts a crossover between the discreet layered superconductivity and the anisotropic three-dimensional superconductivity.
Based on our numerical results we have considered a symmetric stellar setting, as shown in Fig. 10, and have improved the earlier estimates of the stress tensor in lasagna, showing a possibility of shattering the crust by a combined effect of proton superconductivity, magnetic induction and the pulsar spin lag.

\section*{Acknowledgements}
X.V. warmly acknowledges useful discussions with B. K. Sharma and M. Centelles.
X.V. acknowledges partial support from Grants No. PID2023-147112NB-C22 and No. CEX2019-000918-M (through the ``Unit of Excellence Mar\'{\i}a de
Maeztu 2020-2023'' award to ICCUB) from the Spanish MCIN/AEI/10.13039/501100011033.
D.K. acknowledges partial support in the initial stage of the work from Center of Excellence “Center of Photonics” and RSF project 20-12-00268.

\appendix
\section{The energy density functional and properties of uniform nuclear matter}
In this work we use the Skyrme mean-field model introduced by Lim and Holt \cite{LimHolt2017}, which is based on the effective field theory and constrained by the ground-state energies of doubly magic nuclei and observations of neutron stars.
In the Skyrme mean-field approach, an extended Thomas-Fermi approximation is used, where the quantum kinetic energy density is replaced by the semiclassical one \cite{brack85}, which at $\hbar^2$ order is expressed by a density functional that includes second order derivatives of the neutron and proton densities.
Within this approach the energy density for this kind of interactions reads:
\begin{widetext}
\begin{eqnarray}
\varepsilon = \frac{\hbar^2}{2m}(f_n\tau_n+ f_p\tau_p) + \frac{t_0}{4}[(x_0+2)n^2 -(2x_0+1)(n_n^2+n_p^2)] \nonumber \\
- \frac{1}{32}[t_2(2+x_2)-3t1(2+x_1)](\nabla n)^2  - \frac{1}{32}[3t_1(2x_1+1)+t2(2x_2+1)]
[(\nabla n_n)^2+(\nabla n_p)^2] \nonumber \\
+\frac{t_3 n^{\alpha_1}}{24}[(x_3+2)n^2 -(2x_3+1)(n_n^2+n_p^2)]
+ \frac{t_4 n^{\alpha_2}}{24}[(x_4+2)n^2 -(2x_4+1)(n_n^2+n_p^2)] \nonumber \\
-\frac{W_0}{2} (n {\bf \nabla}\cdot{\bf J} + n_n {\bf \nabla}\cdot{\bf J_n} + n_p {\bf \nabla}\cdot{\bf J_p})
\label{eqA11}
\end{eqnarray}
where  $f_q$ ($q=n,p$) is the scaled inverse of the nucleon \emph{effective mass} for each nucleon species, which is defined as
\begin{equation}
f_q = \frac{m}{\tilde{m_q}} = 1 + \frac{m}{4\hbar^2}\{[t_1(x_1+2) + t_2(x_2+2)]n
+ [t_2(2x_2+1) - t_1(2x_1+1)]n_q \}.
\label{eqA12}
\end{equation}
Here, $\tau_q$ and ${\bf J}_q$ are the $\hbar^2-$order extended Thomas-Fermi kinetic energy density and the spin densities, respectively, which read
\begin{equation}
\tau_q  =  \frac{3}{5}(3\pi^2)^{2/3} n_q^{5/3} + \frac{1}{36}\frac{({\bf \nabla}n_q)^2}{n_q}
- \frac{1}{3} \frac{{\bf \nabla}f_q{\bf \nabla}n_q}{f_q} - \frac{1}{12} n_q\frac{({\bf \nabla}f_q)^2}{f^2_q}
+ \frac{({\bf J}_q)^2}{2 n_q} \quad \text{and} \quad
{\bf J}_q = \frac{m}{\hbar^2}W_0 \frac{n_q}{f_q} ({\bf \nabla}n + {\bf \nabla}n_q)
\label{eqA13}
\end{equation}
\end{widetext}
In liquid drop model calculations the relevant part of the energy density functional (\ref{eqA11}) is the bulk part, which we write in a dimensionless
form by scaling the energies and densities by the Fermi energy $\varepsilon_0$ and the saturation density, respectively.
Mapping of the operator form of the interaction energy from Eqs. (3)-(4) of \cite{LimHolt2017} to the energy per baryon $\varepsilon$ in uniform nuclear matter can be done using the formula given in Eq. (1) of \cite{Dutra2012}, leading to:
\begin{widetext}
\begin{eqnarray}
 \nonumber && \varepsilon(n,x)=\frac{3}{5}\varepsilon_0(2\eta)^{\frac{2}{3}}\left\{  \left[x^{\frac{5}{3}}+(1-x)^{\frac{5}{3}}\right](1+\frac{\tilde{a}}{8}\eta) + \left[x^{\frac{8}{3}}+(1-x)^{\frac{8}{3}}\right]\frac{\tilde{b}}{4}\eta  \right\}         \\
 \nonumber && +\frac{t_0n_0}{4}\eta  \left\{ x_0+2-(2x_0+1)\left[x^2+(1-x)^2\right] \right\}\\
 \nonumber && +\frac{t_3n_0^{\alpha_1+1}}{24}\eta^{\alpha_1+1}  \left\{ x_3+2-(2x_3+1)\left[x^2+(1-x)^2\right] \right\}\\
 \label{def_exn} && +\frac{t_4n_0^{\alpha_2+1}}{24}\eta^{\alpha_2+1}  \left\{ x_4+2-(2x_4+1)\left[x^2+(1-x)^2\right] \right\}  ,
\end{eqnarray}
\end{widetext}
where $\eta\equiv n/n_0$,
\begin{eqnarray}
 \label{def_a} && \tilde{a}=\frac{2m_p}{\hbar^2}n_0\left[t_1(x_1+2)+t_2(x_2+2)\right],\\
 \label{def_b} && \tilde{b}=\frac{2m_p}{\hbar^2}n_0\left[t_2(2x_2+1)-t_1(2x_1+1)\right],
\end{eqnarray}
and the numerical values of the parameters are given in Table \ref{table1}.
\begin{table}
\begin{tabular}{|c|c|}
  \hline
  $t_0\;[{\rm MeV\,fm}^{3}]$ & -1803.2928 \\
  \hline
  $t_1\;[{\rm MeV\,fm}^{5}]$ &  301.8208 \\
  \hline
  $t_2\;[{\rm MeV\,fm}^{5}]$ &  -273.2827 \\
  \hline
  $t_3\;[{\rm MeV\,fm}^{3(\alpha_1+1)}]$ &  12783.8619 \\
  \hline
  $t_4\;[{\rm MeV\,fm}^{3(\alpha_2+1)}]$ &  564.1049 \\
  \hline
  $x_0$ &  0.4430 \\
  \hline
  $x_1$ &  -0.3622 \\
  \hline
  $x_2$ &  -0.4105 \\
  \hline
  $x_3$ &  0.6545 \\
  \hline
  $x_4$ &  -11.3160 \\
  \hline
  $\alpha_1$ &  1/3 \\
  \hline
  $\alpha_2$ &  1 \\
  \hline
  $n_0\;[{\rm fm}^{-3}]$ & 0.1561 \\
  \hline
  $P_{\rm nuc}\;[{\rm MeV\,fm}^{-3}]$ & $3\times10^{-6}$ \\
  \hline
  $B\;[{\rm MeV}]$ & -15.91 \\
  \hline
  $K\;[{\rm MeV}]$ & 239.3 \\
  \hline
  $S\;[{\rm MeV}]$ & 31.44 \\
  \hline
  $L\;[{\rm MeV}]$ & 42.07 \\
  \hline
\end{tabular}
\caption{\label{table1} Nuclear interaction parameters are given by $t_{0,1,2,3,4}$, $x_{0,1,2,3,4}$, $\alpha_{1,2}$ \cite{LimHolt2017}.
The empirical nuclear parameters $B$, $K$, $S$, $L$ are calculated from Eqs. (\ref{def_B})-(\ref{def_K}).
The nuclear symmetric saturation density $n_0$ is fine-tuned so that the empirical condition for the pressure, Eq. (\ref{PnucSaturation}), is satisfied to a good precision.}
\end{table}
The critical density at which the uniform nuclear matter becomes unstable with respect to hydrostatic perturbations is
\begin{equation}\label{def_eta_uni_star}
\eta_{\rm uni}^{*}\equiv n_t/n_0,
\end{equation}
with $n_t$ given in Eq. (30) of \cite{LimHolt2017}.

The nuclear saturation condition expresses the fact that the symmetric nuclear matter at saturation is pressureless:
\begin{equation}\label{PnucSaturation}
P_{\rm nuc}(n=n_0,x=1/2)=0.
\end{equation}
The nuclear binding energy per nucleon reads
\begin{equation}\label{def_B}
B=\varepsilon(n=n_0,x=1/2).
\end{equation}
The nuclear symmetry energy is defined as
\begin{equation}\label{def_S}
S=\varepsilon(n=n_0,x=0)-\varepsilon(n=n_0,x=\frac{1}{2})
\end{equation}
and its slope parameter
\begin{equation}\label{def_L}
L=\frac{3}{8}n_0\partial^3_{nxx}\varepsilon|_{n=n_0,\,x=1/2}.
\end{equation}
The incompressibility reads
\begin{equation}\label{def_K}
K=9\partial_{n}P_{\rm nuc}|_{n=n_0,\,x=1/2}.
\end{equation}
The parameters $K$ and $S$ have been constrained by the experiment: the expected range of $K$ is $230\pm30$ MeV and the expected range of $S$ is $32\pm2$ MeV \cite{GonzalezBoquera2019}.
However, the slope parameter $L$ is not well constrained as various models predict different values of $L$.
For instance, the Skyrme interactions favored in \cite{Dutra2012} predict $L$ in the range between 53.04 and and 61.45 MeV; the chiral effective field theory predictions reported in \cite{GramsEtal2022,GramsEtalEPJA2022} span the range between 36.5 and 69.0 MeV.
A recent review is given in \cite{Vinas2014}.

\section{Nucleus phases in liquid drop model}
A complete set of variables of the compressible liquid drop model for the nucleus phases, which include 1N, 2N, 3N phases are
\begin{equation}\label{Nvariables}
  \{n_b,\,n,\,x,\,r_N,\,n_{no}\}.
\end{equation}
In the practical problem, the average baryon density $n_b$ is fixed.
Thus, the unknown variables are
\begin{equation}\label{NvariablesPract}
  \{n,\,x,\,r_N,\,n_{no}\}.
\end{equation}
Their definition is given below.
\subsection{Energy density}
The protons are uniformly distributed within the nucleus with volume $V_N$ with the number density
\begin{equation}\label{def_np}
  n_p=\frac{N_p}{V_N}.
\end{equation}
Here,
\begin{equation}\label{def_VN}
  V_N=\left\{ \begin{array}{c}
                ({4}/{3}) \pi r_N^3,\quad (3N) \\
                \pi r_N^2\times L,\quad (2N) \\
                2r_N\times L^2,\quad (1N)
              \end{array}
   \right.
\end{equation}
where $r_N$ is the radius of the nucleus, $L\rightarrow\infty$ is the length of the rod-like nucleus and $L^2\rightarrow\infty$ is the area of the slab-like nucleus, see Fig. 11 (b).
The limit $L\rightarrow\infty$ should be understood as large enough $L$ so that the Coulomb energy per unit length or area of nuclei and bubbles does not depend on $L$.

The nucleus volume fraction is
\begin{equation}\label{def_u}
  u=\frac{V_N}{V_c}=\left(\frac{r_N}{r_c}\right)^d,
\end{equation}
where $r_c$ is the radius of the unit cell and $d=3$ for 3N, $d=2$ for 2N and $d=1$ for 1N.

Neutrons, however, are distributed within the entire cell, both inside and outside the nucleus.
Moreover, there are neutrons adsorbed on the surface of the nucleus or the bubble.
The adsorbed neutrons are associated with both the neutron skin and the density curvature, \emph{i.e.} the region where the second spatial derivative of the nucleon number density with respect to the coordinate normal to the nucleus surface is nonzero.
Thus, the total number of neutrons in the cell can be split into three parts:
\begin{equation}\label{def_Nss}
  N_n=N_{ni}+N_{surf}+N_{no},
\end{equation}
which corresponds to neutrons inside the nucleus, at the surface and outside the nucleus.
In this paper, we neglect the neutron skin, $N_{surf}=0$.

The \emph{local} number density of neutrons inside the nucleus is
\begin{equation}\label{def_nni}
  n_{ni}=\frac{N_{ni}}{V_N}=(1-x)n,
\end{equation}
where we introduce the standard definition of the proton fraction,
\begin{equation}\label{def_x}
  x=\frac{N_{p}}{N_{p}+N_{ni}}
\end{equation}
and of the baryon density inside the nucleus
\begin{equation}\label{def_n}
  n=\frac{N_{p}+N_{ni}}{V_N}.
\end{equation}
The \emph{local} number density of neutrons outside the nucleus is
\begin{equation}\label{def_nno}
  n_{no}=\frac{N_{no}}{V_c-V_N}.
\end{equation}

The \emph{cell-averaged} number density of baryons can be written as
\begin{equation}\label{def_nbA}
  n_{b}=un+(1-u)n_{no}.
\end{equation}
As we have defined the necessary ingredients, we turn to defining the total energy density
\begin{equation}\label{def_wtot}
  w_{\mathrm{tot}}=\frac{E_{\mathrm {tot}}}{V_c}.
\end{equation}

In liquid drop model, the total energy density includes contributions to the energy per baryon, which we choose to express by the following 6 terms:
\begin{eqnarray}\label{wtot}
&&w_{\mathrm{tot}} \\
\nonumber && =w_{\mathrm{nuc}} + w_{\mathrm{p.surf}} + w_{\mathrm{C+L}} + w_{\mathrm{no}} + w_{\mathrm{e}} + w_{\mathrm{curv}}.
\end{eqnarray}
where $w_{\mathrm{nuc}}$ is the bulk (uniform) energy density of the nuclear cluster.
The terms  $w_{\mathrm{p.surf}}$ and $w_{\mathrm{curv}}$ are the nuclear surface contributions, which represent a \emph{planar surface} contribution and a \emph{curvature} correction.
The contribution from the dripped neutron liquid is given by $w_{\mathrm{no}}$ and the kinetic energy of the electron background by $w_{\mathrm{e}}$.
The Coulomb energy term $w_{\mathrm{C+L}}$ collects the self-energy of protons in the nuclear cluster and of electrons distributed in the WS cell as well as the proton-electron lattice energy.

The rest mass contribution due to baryons and the strong interaction contribution associated with the uniform matter inside the nucleus reads
\begin{equation}\label{def_wnuc}
  w_{\rm nuc}=un\left[\left(1-x\right)m_n + xm_p\right]c^2 + un\varepsilon(n,x),
\end{equation}
where $m_p$ and $m_n$ are the proton and neutron rest masses, correspondingly, $c$ is the speed of light and $\varepsilon$ is the energy per baryon specified in Eq. (\ref{eqA11}).

The planar surface energy density $w_{\rm p.surf}$ is proportional to the surface area of the nucleus and therefore has the form
\begin{equation}\label{def_ws}
  w_{\rm p.surf}=\frac{ud}{r_N}\sigma_s(x),
\end{equation}
where $\sigma_s(x)$ is a function with dimension MeV fm$^{-2}$ and is computed by the method described in \cite{Vinas1998} with the results presented in Appendix D.
The Coulomb energy density including the nucleus self-energy and the lattice energy is \cite{RavenhallEtAl1983,PethickRavenhall1995}
\begin{equation}\label{def_wCoul}
  w_{\rm C+L}=2\pi(enxr_N)^2uf_d(u),
\end{equation}
where
\begin{equation}\label{def_fdu}
  f_d(u)=\frac{1}{d+2}\left[\frac{2}{d-2}\left(1-\frac{du^{1-2/d}}{2}\right) + u\right],
\end{equation}
with
\begin{eqnarray}
\nonumber && f_3(u)=\frac{1}{5}(2-3u^{1/3}+u), \\
\nonumber && f_2(u)=\frac{1}{4}\left(\ln\frac{1}{u}-1+u\right), \\
\nonumber && f_1(u)=\frac{1}{3}\left(\frac{1}{u}-2+u\right).
\end{eqnarray}
Equation (\ref{def_wCoul}) describes a sum of electrostatic energy of a single cluster plus energy of interaction of the cluster with the lattice of other cluster plus the correction due to the finite size of the cluster \cite{PethickRavenhall1995}.
The energy density of the dripped neutrons $w_{\mathrm{no}}$ is
\begin{equation}\label{def_wno}
  w_{\mathrm{no}}=\left(1-u\right)n_{no}\left[m_nc^2 + \varepsilon(n_{no},0)\right],
\end{equation}
the electron energy density $w_{\rm e}$ is
\begin{equation}\label{def_we}
  w_{\rm e}=\frac{3}{4}\hbar c(3\pi^2)^{1/3} (unx)^{4/3},
\end{equation}
where we have used the electrical neutrality condition in the unit cell, so the electron number density $n_e$ reads
\begin{equation}\label{def_ne}
n_e=unx.
\end{equation}

The energy density associated with curvature of the thin surface of nuclear matter inside the nucleus is proportional to both the surface area of the nucleus and to the mean principal curvature of the surface of the nucleus, which is $2/r_N$ for nuclear clusters of 3N, $1/r_N$ for those of 2N and 0 for 1N.
Thus, it has the form
\begin{equation}\label{def_wbend}
  w_{\rm curv}=\frac{ud(d-1)}{r_N^2}\sigma_c,
\end{equation}
where $\sigma_c=\sigma_c(x)$ is a parameter with dimension MeV fm$^{-1}$ and can be found using the method of \cite{Vinas1998}.

\subsection{Variational equations}
The variational equations have the form
\begin{equation}\label{def_VarEq}
  \frac{\partial w_{\rm {tot}}}{\partial y}=0,
\end{equation}
where $y$ is either of the variables from the set $\{n,\,x,\,n_{no},\,r_N,\,u\}$.
In order to reveal the physical significance of the variational equations we shall discuss in detail each of the equations.

\subsection{Optimum size of nucleus}
The extremum
\begin{equation}\label{var_eq_rN}
  \left.\frac{\partial(w_{\rm tot})}{\partial r_N}\right|_{n,x,n_{no},u}=0
\end{equation}
with respect to the nucleus size $r_N$ at fixed $\{n,\,x,\,n_{no},\,u\}$ implies that the energy per baryon is minimized by nucleus size while the number densities are fixed.
Application of the derivative to Eq. (\ref{def_wtot}) yields the result known as the nuclear virial theorem with a correction due to the principal curvature of the thin surface of the dense (nuclear) phase of the inner crust matter:
\begin{equation}\label{ws_2wCL}
  w_{\rm p.surf} + 2w_{\rm curv}=2w_{\rm C+L}.
\end{equation}
Notice that in Eq. (\ref{ws_2wCL}) there are no contributions from neutron skin even if the corresponding energy density were included into $w_{\rm {tot}}$.
From Eq. (\ref{ws_2wCL}) we find the equation that determines $r_N$:
\begin{equation}\label{rNequation}
  r_N^4 - 4qr_N - 3r = 0,
\end{equation}
where
\begin{eqnarray}
\label{qpoly} q = \frac{1}{4}\frac{\sigma_s(x)d}{4\pi(enx)^2f_d(u)}, \\
\label{rpoly} r = \frac{1}{3}\frac{2d(d-1)\sigma_c(x)}{4\pi(enx)^2f_d(u)}.
\end{eqnarray}
The solution to Eq. (\ref{rNequation}) can be readily found using the well-known method \cite{Cardano1545} resulting in
\begin{equation}\label{solution_rN}
  r_N=\sqrt{\frac{p}{2}}\left[1 + \sqrt{q\left(\frac{2}{p}\right)^{3/2}-1}\right],
\end{equation}
where
\begin{equation}\label{p}
  p=\left[q^2 + \sqrt{q^4+r^3}\right]^{1/3} - \left[\sqrt{q^4+r^3}-q^2\right]^{1/3}.
\end{equation}
In case when the curvature contribution is neglected one obtains a simple result,
\begin{equation}\label{solution_rN_noCurvature}
  r_N=\left[\frac{\sigma_s(x) d}{4\pi(enx)^2f_d(u)}\right]^{1/3}. \quad(\sigma_c=0)
\end{equation}

\subsection{Chemical equilibrium}
The extremum
\begin{equation}\label{var_eq_x}
  \left.\frac{\partial(w_{\rm tot})}{\partial x}\right|_{n,n_{no},r_N,u}=0
\end{equation}
implies that the isospin of nucleus is in equilibrium with respect to weak interactions:
\begin{eqnarray}
 &&  \label{betaEquilCond}
  \mu_{e}+(m_p-m_n)c^2 = -\left.\frac{\partial \varepsilon}{\partial x}\right|_{n} \\
\nonumber  &&  - \frac{1}{un}\left.\frac{\partial (w_{\mathrm{p.surf}} + w_{\mathrm{C+L}} + w_{\mathrm{curv}})}{\partial x}\right|_{n,n_{no},r_N,u}.
\end{eqnarray}
Notice that, again, there are no contributions from neutron skin even if the corresponding energy density were included into $w_{\rm {tot}}$.
Here, the electron chemical potential $\mu_{e}$ is:
\begin{equation}\label{def_mu_cluster}
  \mu_{e}=\frac{1}{un}\left.\frac{\partial w_{\mathrm{e}}}{\partial x}\right|_{n,n_{no},r_N,u}=\hbar c(3\pi^2uxn)^{\frac{1}{3}}.
\end{equation}
The chemical equilibrium condition can be written as
\begin{equation}\label{betaEquilCond_form}
  \mu_{e}=\mu_{ni} - \mu_{pi},
\end{equation}
where the neutron and the proton chemical potentials are given below by Eqs. (\ref{Muni}) and (\ref{Mupi}), respectively.

\subsection{Continuity of neutron chemical potential across the nucleus interface}
The extremum
\begin{equation}\label{var_nni}
  \left.\frac{\partial(w_{\rm tot})}{\partial n_{ni}}\right|_{N_{n},n_{pi},r_N,u}=0
\end{equation}
implies that the total energy density is minimized by choosing the optimal number of neutrons inside and outside of the nucleus while keeping their total number fixed as well as fixing the proton number, the size of the nucleus and that of the unit cell.
As a result, the energy to add a neutron to the nucleus is equal to the energy to add a neutron to the dripped neutrons, $\mu_{ni}=\mu_{no}$.

To obtain the explicit form of this condition we note that fixing of $N_{n}$, $n_{pi}$, $r_N$, $u$ induces the following relations:
\begin{eqnarray}
\label{deltax_deltan}&& n\delta x + x\delta n=0,\\
\label{deltanni_deltanno}&& u\delta n_{ni} + (1-u)\delta n_{no}=0.
\end{eqnarray}
The partial derivative reads
\begin{eqnarray}\label{partial_nni_initial}
 &&  \left.\frac{\partial}{\partial n_{ni}}\right|_{N_{n},n_{pi},r_N,u}= \\
  \nonumber && \left.\frac{\partial}{\partial n}\right|_{x,n_{no},r_N,u} + \frac{\delta n}{\delta x}\left.\frac{\partial}{\partial x}\right|_{n,n_{no},r_N,u} \\
  \nonumber && + \frac{\delta n_{no}}{\delta n_{ni}}\left.\frac{\partial}{\partial n_{no}}\right|_{n,x,r_N,u}.
\end{eqnarray}
Using Eqs. (\ref{deltax_deltan}) and (\ref{deltanni_deltanno}), we find
\begin{eqnarray}\label{partial_nni}
 &&  \left.\frac{\partial}{\partial n_{ni}}\right|_{N_{n},n_{pi},r_N,u}= \\
  \nonumber && \left.\frac{\partial}{\partial n}\right|_{x,\,n_{no},r_N,u} - \frac{x}{n}\left.\frac{\partial}{\partial x}\right|_{n,n_{no},r_N,u} \\
  \nonumber && - \frac{u}{1-u}\left.\frac{\partial}{\partial n_{no}}\right|_{n,x,r_N,u}.
\end{eqnarray}
Application of these partial derivatives in Eq. (\ref{var_nni}) yields the promised result,
\begin{equation}\label{continuityMu}
  \mu_{ni}=\mu_{no},
\end{equation}
where
\begin{eqnarray}
\nonumber   \mu_{ni}=m_nc^2 + \left(\left.\frac{\partial}{\partial n}\right|_{x} - \frac{x}{n}\left.\frac{\partial}{\partial x}\right|_{n}\right)\left[n\varepsilon(n,x)\right] \\
\label{Muni}   - \frac{x}{un} \left.\frac{\partial  (w_{\rm p.surf}+w_{\rm curv})}{\partial x}\right|_{r_N,u},
\end{eqnarray}
and
\begin{eqnarray}
\nonumber  \mu_{no}=\frac{1}{1-u}\left.\frac{\partial  w_{\mathrm{no}}}{\partial n_{no}}\right|_{u} \\
\label{Muno} = m_nc^2 + \frac{\partial }{\partial n_{no}}[n_{no}\varepsilon(n_{no},0)].
\end{eqnarray}
Notice that Eqs. (\ref{Muni}) and (\ref{Muno}) do not contain the contribution from $w_{\rm C+L}$ by virtue of Eq. (\ref{deltax_deltan}).
Moreover, there are no contributions from neutron skin even if the corresponding energy density were included into $w_{\rm {tot}}$.
The derivatives within $\mu_{ni}$ are calculated as following:
\begin{eqnarray}
\label{dws_dx} && -\frac{x}{un}\frac{\partial  w_{\rm p.surf}}{\partial x}=-\frac{xd}{nr_N}\frac{\partial  \sigma_s(x)}{\partial x},\\
\label{dwC_dx} && -\frac{x}{un}\frac{\partial  w_{\rm curv}}{\partial x}=-\frac{xd(d-1)}{nr_N^2}\frac{\partial  \sigma_c(x)}{\partial x}.
\end{eqnarray}

By combining Eqs. (\ref{betaEquilCond}), (\ref{betaEquilCond_form}) and (\ref{Muni}) we find the proton chemical potential inside the nucleus including the surface and Coulomb corrections:
\begin{eqnarray}
\nonumber   &&\mu_{pi}=m_pc^2 + \left(\left.\frac{\partial}{\partial n}\right|_{x} + \frac{1-x}{n}\left.\frac{\partial}{\partial x}\right|_{n}\right)\left[n\varepsilon(n,x)\right] \\
\nonumber  &&+ \frac{1-x}{un} \left.\frac{\partial  (w_{\rm p.surf}+w_{\rm curv})}{\partial x}\right|_{r_N,u} \\
\label{Mupi}&&+ \frac{1}{un} \left.\frac{\partial  (w_{\rm C+L})}{\partial x}\right|_{n,r_N,u}.
\end{eqnarray}

The proton chemical potential outside the nucleus can be found directly from the bulk properties of the pure neutron matter:
\begin{equation}
\label{Mupo}  \mu_{po}=\left.\left[\left(\left.\frac{\partial}{\partial n}\right|_{x} + \frac{1}{n}\left.\frac{\partial}{\partial x}\right|_{n}\right)\left[n\varepsilon(n,x)\right]\right]\right|_{n=n_{no},\,x=0}.
\end{equation}

\subsection{Continuity of pressure across the nucleus interface}
The extremum
\begin{equation}\label{var_u}
  \left.\frac{\partial(w_{\rm tot})}{\partial u}\right|_{N_{n},N_{no},x,V_c}=0
\end{equation}
implies that the energy density is minimized by choosing the optimal size of the nucleus while keeping fixed the cell size, the number of neutrons, both inside and outside the nucleus, and the proton fraction.
As a result, the pressure in the nucleus is equal to the pressure in the dripped neutrons, $P_{i}=P_{o}$.

Fixing $N_{n}$, $N_{no}$, $x$ and $V_c$ induces the following relations:
\begin{eqnarray}
\label{deltau_deltan} &&  n\delta u + u\delta n=0,\\
\label{deltanno_deltau}&&  (1-u)\delta n_{no} - n_{no}\delta u=0.
\end{eqnarray}
The partial derivative is reads
\begin{eqnarray}\label{partial_u_initial}
 &&  \left.\frac{\partial}{\partial u}\right|_{N_{n},N_{no},x,V_c}= \\
  \nonumber && \left.\frac{\partial}{\partial u}\right|_{n,x,n_{no},V_c} + \frac{\delta n}{\delta u}\left.\frac{\partial}{\partial n}\right|_{n_{no},x,r_N,u} \\
  \nonumber && + \frac{\delta n_{no}}{\delta u}\left.\frac{\partial}{\partial n_{no}}\right|_{n,x,r_N,u}.
\end{eqnarray}
Using Eqs. (\ref{deltau_deltan}) and (\ref{deltanno_deltau}), we find
\begin{eqnarray}\label{partial_u}
 &&  \left.\frac{\partial}{\partial u}\right|_{N_{n},N_{no},x,V_c}= \\
  \nonumber && \left.\frac{\partial}{\partial u}\right|_{n,x,n_{no},V_c} - \frac{n}{u}\left.\frac{\partial}{\partial n}\right|_{n_{no},x,r_N,u} \\
  \nonumber && + \frac{n_{no}}{1-u}\left.\frac{\partial}{\partial n_{no}}\right|_{n,x,r_N,u}.
\end{eqnarray}
Application of these partial derivatives in Eq. (\ref{var_u}) yields the promised result:
\begin{equation}\label{continuityP}
  P_{i}=P_{o},
\end{equation}
where
\begin{eqnarray}
\nonumber &&  P_{i} = n^2\left.\frac{\partial \varepsilon}{\partial n}\right|_{x} - \left(\left.\frac{\partial}{\partial u}\right|_{n,x,n_{no},V_c} - \frac{n}{u}\left.\frac{\partial}{\partial n}\right|_{n_{no},r_N,u}\right)\\
\label{Pi} && \times(w_{\rm p.surf} + w_{\rm C+L}+w_{\rm curv}),
\end{eqnarray}
and
\begin{equation}
\label{Po} P_{o} = \frac{n_{no}}{1-u}\left.\frac{\partial w_{\mathrm{no}}}{\partial n_{no}}\right|_{u} - \frac{w_{\mathrm{no}}}{1-u} \\
= n_{no}^2\frac{\partial \varepsilon(n_{no},0)}{\partial n_{no}}.
\end{equation}
Notice that there are no contributions from neutron skin even if the corresponding energy density were included into $w_{\rm {tot}}$.
With the help of Eq. (\ref{deltau_deltan}) and the relations
\begin{eqnarray}
&& \left.\partial(u/r_N)/\partial u\right|_{V_c}=(d-1)/r_Nd,\\
&& \left.\partial(u/r_N^2)/\partial u\right|_{V_c}=(d-2)/r_N^2d,\\
&& \left.\partial(r_N^2)/\partial u\right|_{V_c}=2r_N^2/ud,
\end{eqnarray}
the derivatives within $P_{i}$ are calculated as following:
\begin{eqnarray}
\label{dws_du}&&\left.\frac{\partial w_{\rm p.surf}}{\partial u}\right|_{N_{n},N_{no},x,V_c}=\frac{d-1}{r_N}\sigma_s(x),\\
\label{dwc_du}&&\left.\frac{\partial w_{\rm curv}}{\partial u}\right|_{N_{n},N_{no},x,V_c}=\frac{(d-2)(d-1)}{r_N^2}\sigma_c(x),
\end{eqnarray}
and
\begin{eqnarray}
  \nonumber  \left(\left.\frac{\partial}{\partial u}\right|_{n,x,n_{no},V_c} - \frac{n}{u}\left.\frac{\partial}{\partial n}\right|_{n_{no},r_N,u}\right)w_{\rm C+L}\\
  \label{dwCL_du} =2\pi (e x n r_N)^2 (u-1)g_d,
\end{eqnarray}
where $g_3=2/15$, $g_2=1/4$ and $g_1=2/3$.

\section{Bubble phases in liquid drop model}
A complete set of independent variables for the bubble phases 1B, 2B, 3B are
\begin{equation}\label{Nvariables_BUB}
  \{n_b,\,n,\,x,\,r_B,\,n_{no}\},
\end{equation}
where the baryon density and the proton fraction outside the bubble $n$ and $x$, correspondingly, the bubble radius $r_B$.
As in the case with nuclear clusters, in the practical problem the unknown variables are:
\begin{equation}\label{PractNvariables_BUB}
  \{n,\,x,\,r_B,\,n_{no}\},
\end{equation}

\subsection{Energy density}
The protons are uniformly distributed outside the bubble with volume $V_B$ with the number density
\begin{equation}\label{def_np_BUB}
  n_p=\frac{N_p}{V_c-V_B}.
\end{equation}
Here,
\begin{equation}\label{def_VN_BUB}
  V_B=\left\{ \begin{array}{c}
                ({4}/{3}) \pi r_B^3,\quad (3B) \\
                \pi r_B^2\times L,\quad (2B) \\
                2r_B\times L^2,\quad (1B)
              \end{array}
   \right.
\end{equation}
where $r_B$ is the radius of the bubble, $L\rightarrow\infty$ is the length of the rod-like bubble and $L^2\rightarrow\infty$ is the area of the slab-like bubble.

The bubble volume fraction is
\begin{equation}\label{def_u_BUB}
  u^{\rm bub}=\frac{V_B}{V_c}=\left(\frac{r_B}{r_c}\right)^d,
\end{equation}
where $r_c$ is the radius of the unit cell, $d=3$ for 3B and $d=2$ for 2B.

The total number of neutrons in the cell can be split into three parts:
\begin{equation}\label{def_Nss_BUB}
  N_n=N_{ni}+N_{surf}+N_{no},
\end{equation}
which corresponds to neutrons outside the bubble, at the surface and inside the bubble.
Again, we assume $N_{surf}=0$.
We define the following neutron number densities.

The \emph{local} number density of neutrons outside the bubble is
\begin{equation}\label{def_nni_BUB}
  n_{ni}=\frac{N_{ni}}{V_c-V_B}=(1-x)n,
\end{equation}
where we introduce the standard definition of the proton fraction,
\begin{equation}\label{def_x_BUB}
  x=\frac{N_{p}}{N_{p}+N_{ni}}
\end{equation}
and of the baryon density outside the bubble
\begin{equation}\label{def_n_BUB}
  n=\frac{N_{p}+N_{ni}}{V_c-V_B}.
\end{equation}
The \emph{local} number density of neutrons inside the bubble is
\begin{equation}\label{def_nno_BUB}
  n_{no}=\frac{N_{no}}{V_B}.
\end{equation}

Consequently, the \emph{cell-averaged} number density of baryons can be written as
\begin{equation}\label{def_nb_BUB}
  n_{b}=(1-u^{\rm bub})n+u^{\rm bub}n_{no}.
\end{equation}
As we have defined the necessary ingredients, we turn to defining the total energy density
\begin{equation}\label{def_wtot_BUB}
  w_{\mathrm{tot}}^{\rm bub}=\frac{E_{\mathrm {tot}}^{\rm bub}}{V_c}.
\end{equation}

In our model, the total energy density includes 6 terms:
\begin{equation}\label{wtot_BUB}
w_{\mathrm{tot}}^{\rm bub}=w_{\mathrm{nuc}}^{\rm bub} + w_{\mathrm{p.surf}}^{\rm bub} + w_{\mathrm{C+L}}^{\rm bub} + w_{\mathrm{no}}^{\rm bub} + w_{\mathrm{e}}^{\rm bub} + w_{\mathrm{curv}}^{\rm bub}.
\end{equation}
The rest mass contribution due to baryons and the strong interaction contribution associated with the uniform matter outside the bubble reads
\begin{equation}\label{def_wnuc_BUB}
  w_{\rm nuc}^{\rm bub}=(1-u^{\rm bub})n\left[\left(1-x\right)m_n + xm_p\right]c^2 + (1-u^{\rm bub})n\varepsilon(n,x),
\end{equation}
where $\varepsilon$ is the energy per baryon specified in Eq. (\ref{eqA11}).

The surface energy density $w_{\rm p.surf}^{\rm bub}$ is proportional to the surface area of the bubble and therefore has the form
\begin{equation}\label{def_ws_BUB}
  w_{\rm p.surf}^{\rm bub}=\frac{u^{\rm bub}d}{r_B}\sigma_s(x),
\end{equation}
where $\sigma_s(x)$ is the function that has appeared in Eq. (\ref{def_ws}).
The Coulomb energy density including the nucleus self-energy and the lattice energy is \cite{RavenhallEtAl1983,PethickRavenhall1995}
\begin{equation}\label{def_wCoul_BUB}
  w_{\rm C+L}^{\rm bub}=2\pi(enxr_B)^2u^{\rm bub}f_d(u^{\rm bub}),
\end{equation}
where the function $f_d(u)$ is given in Eq. (\ref{def_fdu}).

The energy density of the dripped neutrons $w_{\mathrm{no}}^{\rm bub}$ is
\begin{equation}\label{def_wno_BUB}
  w_{\mathrm{no}}^{\rm bub}=u^{\rm bub}n_{no}\left[m_nc^2 + \varepsilon(n_{no},0)\right],
\end{equation}
the electron energy density $w_{\rm e}$ is
\begin{equation}\label{def_we_BUB}
  w_{\rm e}=\frac{3}{4}\hbar c(3\pi^2)^{1/3} \left[(1-u^{\rm bub})nx\right]^{4/3},
\end{equation}
where we have used the electrical neutrality condition in the unit cell.

The mean principal curvature of the bubble surface is $-2/r_B$ for bubbles of 3B and $-1/r_B$ for bubbles of 2B.
Thus, the energy density associated with curvature of the nucleus surface has the form
\begin{equation}\label{def_wbend_BUB}
  w_{\rm curv}^{\rm bub}=-\frac{u^{\rm bub}d(d-1)}{r_B^2}\sigma_c.
\end{equation}
Notice that the surface curvature contribution for bubbles has the opposite sign as compared with that for nuclei.

\subsection{Variational equations}
The variational equations have the form
\begin{equation}\label{def_VarEq_BUB}
  \frac{\partial w_{\rm {tot}}^{\rm bub}}{\partial y}=0,
\end{equation}
where $y$ is either of the variables from the set $\{n,\,x,\,n_{no},\,r_B,\,u^{\rm bub}\}$.

\subsection{Optimum size of bubble}
The extremum
\begin{equation}\label{var_eq_rN_BUB}
  \left.\frac{\partial(w_{\rm tot}^{\rm bub})}{\partial r_B}\right|_{n,x,n_{no},u^{\rm bub}}=0
\end{equation}
leads to the nuclear virial theorem:
\begin{equation}\label{ws_2wCL_BUB}
  w_{\rm p.surf}^{\rm bub} + 2w_{\rm curv}^{\rm bub}=2w_{\rm C+L}^{\rm bub}.
\end{equation}
From Eq. (\ref{ws_2wCL_BUB}) we find the equation that determines $r_N$:
\begin{equation}\label{rNequation_BUB}
  r_B^4 - 4q^{\rm bub}r_B + 3r^{\rm bub} = 0,
\end{equation}
where
\begin{eqnarray}
\label{qpoly_BUB} q^{\rm bub} = \frac{1}{4}\frac{\sigma_s(x)d}{4\pi(enx)^2f_d(u^{\rm bub})}, \\
\label{rpoly_BUB} r^{\rm bub} = \frac{1}{3}\frac{2d(d-1)\sigma_c(x)}{4\pi(enx)^2f_d(u^{\rm bub})}.
\end{eqnarray}
Notice a difference in signs in Eqs. (\ref{rNequation}) and (\ref{rNequation_BUB}).
The solution to Eq. (\ref{rNequation_BUB}) is
\begin{equation}\label{solution_rN_BUB}
  r_B=\sqrt{\frac{p^{\rm bub}}{2}}\left[1 + \sqrt{q^{\rm bub}\left(\frac{2}{p^{\rm bub}}\right)^{3/2}-1}\right],
\end{equation}
where
\begin{eqnarray}
\nonumber  &&p^{\rm bub}=\left[{q^{\rm bub}}^2 + \sqrt{{q^{\rm bub}}^4-{r^{\rm bub}}^3}\right]^{1/3} \\
\label{p_BUB}  &&+ \left[{q^{\rm bub}}^2-\sqrt{{q^{\rm bub}}^4-{r^{\rm bub}}^3}\right]^{1/3}.
\end{eqnarray}
In case when the curvature contribution is neglected one obtains a simple result,
\begin{equation}\label{solution_rN_noCurvature_BUB}
  r_B=\left[\frac{\sigma_s(x) d}{4\pi(enx)^2f_d(u^{\rm bub})}\right]^{1/3}. \quad(\sigma_c=0)
\end{equation}

\subsection{Chemical equilibrium}
The extremum
\begin{equation}\label{var_eq_x_BUB}
  \left.\frac{\partial(w_{\rm tot}^{\rm bub})}{\partial x}\right|_{n,n_{no},r_B,u^{\rm bub}}=0
\end{equation}
provides the beta-equilibrium condition:
\begin{eqnarray}
 &&  \label{betaEquilCond_BUB}
  \mu_e^{\mathrm{bub}}+(m_p-m_n)c^2 = \\
\nonumber  && -\left.\frac{\partial \varepsilon}{\partial x}\right|_{n} - \frac{1}{(1-u^{\mathrm{bub}})n}\left.\frac{\partial (w_{\rm s}^{\mathrm{bub}}+w_{\rm curv}^{\mathrm{bub}}+w_{\rm C+L}^{\mathrm{bub}})}{\partial x}\right|_{n,n_{no},r_B,u^{\mathrm{bub}}},
\end{eqnarray}
where
\begin{equation}\label{def_mue_BUB}
  \mu_e^{\mathrm{bub}}=\hbar c\left[3\pi^2(1-u^{\mathrm{bub}})xn\right]^{\frac{1}{3}}.
\end{equation}

Again, the chemical equilibrium condition can be written as
\begin{equation}\label{betaEquilCond_form_BUB}
  \mu_{e}^{\mathrm{bub}}=\mu_{ni}^{\mathrm{bub}} - \mu_{pi}^{\mathrm{bub}},
\end{equation}
where the neutron and the proton chemical potentials are given below by Eqs. (\ref{Muni_BUB}) and (\ref{Mupi_BUB}), respectively.

\subsection{Continuity of neutron chemical potential across the bubble interface}
The extremum
\begin{equation}\label{var_nni_BUB}
  \left.\frac{\partial(w_{\rm tot}^{\mathrm{bub}})}{\partial n_{ni}}\right|_{N_{n},n_{pi},r_B,u^{\mathrm{bub}}}=0
\end{equation}
implies that the energy to add a neutron to the nuclear matter outside the bubble is equal to the energy to add a neutron to the dripped neutrons inside the bubble, $\mu_{ni}^{\mathrm{bub}}=\mu_{no}^{\mathrm{bub}}$.

To obtain the explicit form of this condition we note that fixing of $N_{n}$, $n_{pi}$, $r_B$, $u^{\mathrm{bub}}$ induces the following relations:
\begin{eqnarray}
\label{deltax_deltan_bubbles} && n\delta x + x\delta n=0,\\
\label{deltanni_deltanno_bubbles} &&  (1-u^{\mathrm{bub}})\delta n_{ni} + u^{\mathrm{bub}}\delta n_{no}=0.
\end{eqnarray}
The partial derivative reads
\begin{eqnarray}\label{partial_nnibubbles_initial}
 &&  \left.\frac{\partial}{\partial n_{ni}}\right|_{N_{n},\,n_{pi},\,r_B,\,u^{\mathrm{bub}}}= \\
  \nonumber && \left.\frac{\partial}{\partial n}\right|_{x,\,n_{no},\,r_B,\,u^{\mathrm{bub}}} + \frac{\delta x}{\delta n}\left.\frac{\partial}{\partial x}\right|_{n,\,n_{no},\,r_B,\,u^{\mathrm{bub}}} \\
   \nonumber &&+ \frac{\delta n_{no}}{\delta n_{ni}}\left.\frac{\partial}{\partial n_{no}}\right|_{n,\,x,\,r_B,\,u^{\mathrm{bub}}}.
\end{eqnarray}
Using Eqs. (\ref{deltax_deltan_bubbles}) and (\ref{deltanni_deltanno_bubbles}), we find
\begin{eqnarray}\label{partial_nni_BUB}
 &&  \left.\frac{\partial}{\partial n_{ni}}\right|_{N_{n},\,n_{pi},\,r_B,\,u^{\mathrm{bub}}}= \\
  \nonumber && \left.\frac{\partial}{\partial n}\right|_{x,\,n_{no},\,r_B,\,u^{\mathrm{bub}}} - \frac{x}{n}\left.\frac{\partial}{\partial x}\right|_{n,\,n_{no},\,r_B,\,u^{\mathrm{bub}}} \\
  \nonumber && - \frac{1-u^{\mathrm{bub}}}{u^{\mathrm{bub}}}\left.\frac{\partial}{\partial n_{no}}\right|_{n,\,x,\,r_B,\,u^{\mathrm{bub}}}.
\end{eqnarray}
Application of these partial derivatives yields
\begin{equation}\label{continuityMu_BUB}
  \mu_{ni}^{\mathrm{bub}}=\mu_{no}^{\mathrm{bub}},
\end{equation}
where
\begin{eqnarray}
\nonumber   \mu_{ni}^{\mathrm{bub}}=m_nc^2 + \left(\left.\frac{\partial}{\partial n}\right|_{x} - \frac{x}{n}\left.\frac{\partial}{\partial x}\right|_{n}\right)\left[n\varepsilon(n,x)\right] \\
\label{Muni_BUB}   - \frac{x}{(1-u^{\mathrm{bub}})n} \left.\frac{\partial \left( w_{\rm p.surf}^{\mathrm{bub}} + w_{\rm cur}^{\mathrm{bub}}\right)}{\partial x}\right|_{r_B,\,u^{\mathrm{bub}},},
\end{eqnarray}
and
\begin{equation}\label{Muno_BUB}
  \mu_{no}^{\mathrm{bub}}=\frac{1}{u^{\mathrm{bub}}}\left.\frac{\partial  w_{\mathrm{no}}^{\mathrm{bub}}}{\partial n_{no}}\right|_{u^{\mathrm{bub}}}.
\end{equation}

By combining Eqs. (\ref{betaEquilCond_BUB}), (\ref{betaEquilCond_form_BUB}) and (\ref{Muni_BUB}) we find the proton chemical potential outside the bubble including the surface and Coulomb corrections:
\begin{eqnarray}
\nonumber   &&\mu_{pi}^{\mathrm{bub}}=m_pc^2 + \left(\left.\frac{\partial}{\partial n}\right|_{x} + \frac{1-x}{n}\left.\frac{\partial}{\partial x}\right|_{n}\right)\left[n\varepsilon(n,x)\right] \\
\nonumber  &&+ \frac{1-x}{(1-u^{\mathrm{bub}})n} \left.\frac{\partial  (w_{\rm p.surf}^{\rm bub}+w_{\rm curv}^{\rm bub})}{\partial x}\right|_{r_B,u^{\mathrm{bub}}} \\
\label{Mupi_BUB}&&+ \frac{1}{(1-u^{\mathrm{bub}})n} \left.\frac{\partial  (w_{\rm C+L}^{\rm bub})}{\partial x}\right|_{n,r_B,u}.
\end{eqnarray}

\subsection{Continuity of pressure across the bubble interface}
The extremum
\begin{equation}\label{var_u_BUB}
  \left.\frac{\partial(w_{\rm tot}^{\mathrm{bub}})}{\partial u^{\mathrm{bub}}}\right|_{N_{n},N_{no},x,V_c}=0
\end{equation}
implies that the energy density is minimized by choosing the optimal size of the bubble while keeping fixed the cell size, the number of neutrons, both inside and outside the bubble, and the proton fraction.
As a result, the pressure inside the bubble is equal to that outside the bubble, $P_{i}^{\mathrm{bub}}=P_{o}^{\mathrm{bub}}$.

Fixing $N_{n}$, $N_{no}$, $x$ and $V_c$ induces the following relations:
\begin{eqnarray}
\label{deltau_deltan_bubbles} (1-u^{\mathrm{bub}})\delta n - n\delta u^{\mathrm{bub}}=0,\\
\label{deltanno_deltau_bubbles} n_{no} \delta u^{\mathrm{bub}} + u^{\mathrm{bub}}\delta n_{no}=0.
\end{eqnarray}
The partial derivative reads
\begin{eqnarray}\label{partial_ububbles_initial}
 &&  \left.\frac{\partial}{\partial n_{ni}}\right|_{N_{n},\,n_{pi},\,r_B,\,u^{\mathrm{bub}}}=\left.\frac{\partial}{\partial u^{\mathrm{bub}}}\right|_{n,\,x,\,n_{no},\,V_c} \\
  \nonumber &&  + \frac{\delta n}{\delta u^{\mathrm{bub}}}\left.\frac{\partial}{\partial n}\right|_{n_{no},\,x,\,r_B,\,u^{\mathrm{bub}}} + \frac{\delta n_{no}}{\delta u^{\mathrm{bub}}}\left.\frac{\partial}{\partial n_{no}}\right|_{n,\,x,\,r_B,\,u^{\mathrm{bub}}}.
\end{eqnarray}
Using Eqs. (\ref{deltau_deltan_bubbles}) and (\ref{deltanno_deltau_bubbles}), we find
\begin{eqnarray}\label{partial_ububbles}
 &&  \left.\frac{\partial}{\partial u^{\mathrm{bub}}}\right|_{N_{n},\,N_{no},\,x,\,V_c}=\left.\frac{\partial}{\partial u^{\mathrm{bub}}}\right|_{n,\,x,\,n_{no},\,V_c} \\
  \nonumber &&  + \frac{n}{1-u^{\mathrm{bub}}}\left.\frac{\partial}{\partial n}\right|_{n_{no},\,x,\,r_B,\,u^{\mathrm{bub}}} - \frac{n_{no}}{u^{\mathrm{bub}}}\left.\frac{\partial}{\partial n_{no}}\right|_{n,\,x,\,r_B,\,u^{\mathrm{bub}}}.
\end{eqnarray}
Making use of the relation
\begin{equation}\label{relation_bubbles}
\left.\partial[(1-u^{\mathrm{bub}})n]/\partial u^{\mathrm{bub}}\right|_{N_{n},\,N_{no},\,x,\,V_c}=0,
\end{equation}
we apply these partial derivatives and obtain
\begin{equation}\label{continuityP_BUB}
  P_{i}^{\mathrm{bub}}=P_{o}^{\mathrm{bub}},
\end{equation}
where
\begin{eqnarray}
  \nonumber && P_{i}^{\mathrm{bub}} = n^2\left.\frac{\partial \varepsilon}{\partial n}\right|_{x}\\
  \label{Pi_BUB} && + \left(\left.\frac{\partial}{\partial u^{\mathrm{bub}}}\right|_{n,\,x,\,n_{no},\,V_c} + \frac{n}{1-u^{\mathrm{bub}}}\left.\frac{\partial}{\partial n}\right|_{n_{no},\,r_B,\,u^{\mathrm{bub}}}\right)\\
  \nonumber && \times(w_{\rm p.surf}^{\mathrm{bub}}+w_{\rm cur}^{\mathrm{bub}} + w_{\rm C+L}^{\mathrm{bub}}),
\end{eqnarray}
and
\begin{eqnarray}\label{Po_BUB}
  &&P_{o}^{\mathrm{bub}} = \frac{n_{no}}{u^{\mathrm{bub}},n_{sk}^{\mathrm{bub}},n_{dc}^{\mathrm{bub}}}\left.\frac{\partial w_{\mathrm{no}}^{\mathrm{bub}}}{\partial n_{no}}\right|_{u^{\mathrm{bub}}} - \frac{w_{\mathrm{no}}^{\mathrm{bub}}}{u^{\mathrm{bub}}} \\
  \nonumber && =n_{no}^2\frac{\partial \varepsilon(n_{no},0)}{\partial n_{no}}.
\end{eqnarray}
With the help of Eq. (\ref{deltau_deltan}) and the relations
\begin{eqnarray}
&& \left.\partial(u^{\mathrm{bub}}/r_B)/\partial u^{\mathrm{bub}}\right|_{V_c}=(d-1)/r_Bd,\\
&& \left.\partial(u^{\mathrm{bub}}/r_B^2)/\partial u^{\mathrm{bub}}\right|_{V_c}=(d-2)/r_Bd,\\
&& \left.\partial(r_B^2)/\partial u\right|_{V_c}=2r_B^2/u^{\mathrm{bub}}d,
\end{eqnarray}
the derivatives within $P_{i}^{\mathrm{bub}}$ are calculated as following:
\begin{eqnarray}
\label{dws_dububbles}&&\left.\frac{\partial w_{\rm p.surf}^{\mathrm{bub}}}{\partial u}\right|_{N_{n},\,N_{no},\,x,\,V_c}=\frac{d-1}{r_B}\sigma_s(x),\\
\label{dwc_dububbles}&&\left.\frac{\partial w_{\rm cur}^{\mathrm{bub}}}{\partial u}\right|_{N_{n},\,N_{no},\,x,\,V_c}=-\frac{(d-2)(d-1)}{r_B^2}\sigma_c(x),
\end{eqnarray}
and
\begin{eqnarray}
  \nonumber && \left(\left.\frac{\partial}{\partial u}\right|_{n,\,x,\,n_{no},\,V_c} + \frac{n}{1-u^{\mathrm{bub}}}\left.\frac{\partial}{\partial n}\right|_{n_{no},\,r_B,\,u^{\mathrm{bub}}}\right)w_{\rm C+L}^{\mathrm{bub}}\\
  \label{dwCL_dububbles}&&=2\pi (e x n r_B)^2 \\
  \nonumber &&\times\left[\left(\frac{2+d}{d} + \frac{2u^{\mathrm{bub}}}{1-u^{\mathrm{bub}}}\right)f_d(u^{\mathrm{bub}}) + u^{\mathrm{bub}}\partial_{u^{\mathrm{bub}}}f_d(u^{\mathrm{bub}})\right].
\end{eqnarray}

\section{Nuclear surface tension with curvature corrections}
\begin{table}
\begin{tabular}{|c|c|c|c|c|}
  \hline
  $x$ & $\sigma_s\,[{\rm MeV fm}^{-2}]$ & $\partial\sigma_s/\partial x$ & $\sigma_c\,[{\rm MeV fm}^{-1}]$ & $\partial\sigma_c/\partial x$ \\
  \hline
  0.0250 & 1.23$\times10^{-4}$ & 0.01154 & -1.0$\times10^{-4}$ & 0.1258  \\
  \hline
0.0375 & 9.91$\times10^{-4}$ & 0.1304 & 0.001521 & 0.1875 \\
  \hline
0.0500 & 0.003427 & 0.2598 & 0.005366 & 0.4360 \\
  \hline
0.0625 & 0.007467 & 0.3802 & 0.01206 & 0.5526 \\
  \hline
0.0750 & 0.01295 & 0.5053 & 0.01880 & 0.5502 \\
  \hline
0.0875 & 0.02018 & 0.6520 & 0.02606 & 0.6368 \\
  \hline
0.1000 & 0.02934 & 0.8179 & 0.03489 & 0.7670 \\
  \hline
0.1125 & 0.04073 & 1.0079 & 0.04528 & 0.8994 \\
  \hline
0.1250 & 0.05463 & 1.2210 & 0.05746 & 1.0527 \\
  \hline
0.1375 & 0.07135 & 1.4583 & 0.07168 & 1.2267 \\
  \hline
0.1500 & 0.09119 & 1.7198 & 0.08822 & 1.4249 \\
  \hline
0.1625 & 0.1144 & 2.0058 & 0.1074 & 1.6380 \\
  \hline
0.1750 & 0.1414 & 2.3133 & 0.1292 & 1.8672 \\
  \hline
0.1875 & 0.1723 & 2.6371 & 0.1541 & 2.1066 \\
  \hline
0.2000 & 0.2074 & 2.9697 & 0.1819 & 2.3434 \\
  \hline
0.2125 & 0.2466 & 3.3007 & 0.2126 & 2.5542 \\
  \hline
0.2250 & 0.2898 & 3.6179 & 0.2456 & 2.7174 \\
  \hline
0.2375 & 0.3369 & 3.9049 & 0.2803 & 2.8424\\
  \hline
0.2500 & 0.3873 & 4.1527 & 0.3163 & 2.8970 \\
  \hline
0.2625 & 0.4405 & 4.3513 & 0.3525 & 2.8886  \\
  \hline
0.2750 & 0.4959 & 4.4946 & 0.3883 & 2.8267  \\
  \hline
0.2875 & 0.5526 & 4.5808 & 0.4229 & 2.7054  \\
  \hline
0.3000 & 0.6101 & 4.6061 & 0.4558 & 2.5453 \\
  \hline
0.3125 & 0.6675 & 4.5678 & 0.4864 & 2.3485 \\
  \hline
0.3250 & 0.7240 & 4.4627 & 0.5143 & 2.1193 \\
  \hline
0.3375 & 0.7786 & 4.2349 & 0.5386 & 1.6866 \\
  \hline
0.3500 & 0.8294 & 3.8600 & 0.5555 & 0.9633 \\
  \hline
0.3625 & 0.8750 & 3.4700 & 0.5631 & 0.3690 \\
  \hline
0.3750 & 0.9163 & 3.1317 & 0.5657 & 0.04603 \\
  \hline
0.3875 & 0.9534 & 2.8083 & 0.5646 & -0.2078 \\
  \hline
0.4000 & 0.9865 & 2.4923 & 0.5607 & -0.4167 \\
  \hline
0.4125 & 1.0157 & 2.1788 & 0.5543 & -0.5977 \\
  \hline
0.4250 & 1.0410 & 1.8671 & 0.5458 & -0.7558 \\
  \hline
0.4375 & 1.0624 & 1.5563 & 0.5355 & -0.8968\\
  \hline
0.4500 & 1.0799 & 1.2455 & 0.5235 & -1.0245 \\
  \hline
0.4625 & 1.0935 & 0.9344 & 0.5099 & -1.1417\\
  \hline
0.4750 & 1.1033 & 0.6232 & 0.4949 & -1.2502\\
  \hline
0.4875 & 1.1091 & 0.3116 & 0.4787 & -1.3516 \\
  \hline
0.5000 & 1.1110 & -1.2$\times10^{-4}$ & 0.4612 & -1.4472\\
  \hline
\end{tabular}
\caption{\label{table2} The leading (planar surface) contribution $\sigma_s(x)$ and the next-to-leading (principal curvature) correction $\sigma_c(x)$, for the surface tension of a thin surface, and their derivative with respect to the proton fraction $x$, as functions of $x$.
The results obtained on the basis of a unified (self-consistent) description of the bulk and surface nuclear properties by the single Sk$\chi$450 interaction.}
\end{table}
To obtain self-consistently the nuclear surface tension with curvature corrections, let us start considering a neutron-rich nucleus in equilibrium with a liquid of dripped neutrons.
The problem of finding the density profile is formally similar to the one found in hot nucleus in equilibrium with the pure liquid phase.
It is known that in this scenario the variational equations at given temperature and chemical potential possess two different solutions in both quantum \cite{bonche85} and semiclassical descriptions \cite{suraud87}.
One of these solutions represents the nucleus-plus-liquid system and the another one to the pure liquid.
Therefore, it seems natural to define the extensive quantities in the nucleus as the differences of their values in the nucleus-plus-liquid and the pure-liquid system.

To simplify this calculation, we simulate the nucleus as a half-space filled with uniform nuclear matter ($\tilde{z}<0$), while the other half-space
is filled with the pure neutron liquid ($\tilde{z}>0$).
As a consequence, in this geometry the energy per unit area, i.e. the surface tension, of nucleus constrained by the conservation of the neutron and protons numbers reads \cite{Vinas1998}:
\begin{equation}
\sigma_s = \int^{+ \infty}_{- \infty} d\tilde{z} \big[ \varepsilon (\tilde{z}) - \varepsilon_{no} - \mu_n (n_n(\tilde{z}) - n_{no}) - \mu_p n_p (\tilde{z}) \big],
\label{eqA1}
\end{equation}
where $\varepsilon_d$ is the energy density of the neutron liquid and $\varepsilon (\tilde{z})$ is the local energy density of the nucleus plus pure liquid (LG) system, which in the limit $\tilde{z} \to - \infty$ corresponds to the energy density $\varepsilon_0$ of uniform asymmetric nuclear matter in equilibrium with the uniform liquid of dripped neutrons (G).
Here, $n_n(\tilde{z})$ is the total number density of neutrons at the plane $\tilde{z}=\mathrm{const}$, $n_{no}$ is the number density of neutrons of the pure neutron liquid at $\tilde{z} \to + \infty$ and $n_p=n_p(\tilde{z})$ is the number density of the proton component of nuclear matter.

The neutron and proton density profiles as well as the density of dripped neutrons are the solution of the
Euler-Lagrange equations derived from Eq. (\ref{eqA1}):
\begin{equation}
\frac{\delta \varepsilon(\tilde{z})}{\delta n_n} - \mu_n =0, \quad  \frac{\delta \varepsilon(\tilde{z})}{\delta n_p} - \mu_p =0, \quad
\frac{\delta \varepsilon_{no}}{\delta n_{no}} - \mu_n =0,
\label{eqA2}
\end{equation}
where $\mu_n$ and $\mu_p$ are the neutron and proton chemical potentials computed at $\tilde{z} \to - \infty$, i.e. corresponding to the bulk phase
in equilibrium with the dripped neutron liquid, which due to the equilibrium has the same neutron chemical potential $\mu_n$, and $\varepsilon_{no}=\varepsilon(n=n_{no},x=0)$.

The Euler-Lagrange equations (\ref{eqA2}) is a set of coupled second order differential equations, which are solved using the imaginary timestep method and provide self-consistent neutron and proton densities in LG and G phases.
It is worth noting here that the surface tension defined by Eq. (\ref{eqA1}) corresponds to the surface contribution to the thermodynamical potential $\Omega = E - \mu_n N - \mu_p Z= -PV$ in LG phase, which is minimized in thermodynamic equilibrium between LG and G phases \cite{douchin00}.

The surface tension in Eq. (\ref{eqA1}) has been obtained under the assumption of a planar interface.
In order to take into account that the interface is curved, we add an additional curvature correction that takes into account the curvature of the interface \cite{douchin00}.
This corresponds to a Taylor expansion of the surface tension in the curved interface around the planar one.
The lowest order corresponds to the planar surface tension, Eq. (\ref{eqA1}).
The correction term, which is linear in the principal curvature of the surface, is the curvature correction, which reads \cite{Vinas1998,douchin00}:
\begin{equation}
\sigma_c = \int^{+ \infty}_{- \infty} d\tilde{z} [ \varepsilon (\tilde{z}) - \varepsilon_{no} - \mu_n (n_n(\tilde{z}) - n_{no}) - \mu_p n_p(\tilde{z}) ](\tilde{z}-z_p),
\label{eqA7}
\end{equation}
where $z_p$ is the proton surface location, which is defined as $\int^{+ \infty}_{- \infty} d\tilde{z} (\tilde{z} - z_p) n'_p(\tilde{z}) = 0$.

In Table \ref{table2} we display our data for the nuclear surface $\sigma_s$ (in units of ${\rm MeV\,fm}^{-2}$), the curvature $\sigma_c$  (in units of ${\rm MeV\,fm}^{-1}$) tensions and their derivatives with respect to $x$, which are needed to find the solution to the basic equations.

The data for the surface tension may be fitted by analytical functions.
For the planar surface tension, the fit is given in Eq. (\ref{def_sigmaLimHolt}).
For the curvature correction, the fit is
\begin{equation}\label{def_sigmaCfit}
  \sigma_c^{\rm fit}(x)= \frac{\sigma_{c0}}{\sigma_{0}}\sigma_s^{\rm fit}(x) \alpha_c(\beta_c - x).
\end{equation}
For Sk$\chi450$, we find $\sigma_{c0}=0.4612$, $\alpha_c=3.9027$ and $\beta_c=0.7562$.

\end{document}